\documentclass[aps,prd,nobibnotes,twocolumn,superscriptaddress,bibliography]{revtex4-1}

\usepackage{amsfonts}
\usepackage{mathrsfs}
\usepackage{amsmath}
\usepackage{color}
\usepackage{graphicx}
\usepackage{bm}
\usepackage{amssymb}
\usepackage{xspace}
\usepackage{epstopdf}
\usepackage{dcolumn}
\usepackage{longtable}
\usepackage{multirow}
\usepackage{float}
\usepackage{comment}
\usepackage[colorlinks=true, letterpaper=true, pdfstartview=FitV, linkcolor=blue, citecolor=blue, urlcolor=blue]{hyperref}

\providecommand{\tabularnewline}{\\}

\begin{document}
\title{Higher-order Dirac fermions in three dimensions}

\author{Weikang Wu}


\address{Research Laboratory for Quantum Materials, Singapore University of Technology and Design, Singapore 487372, Singapore}

\author{Zhi-Ming Yu}


\address{Research Laboratory for Quantum Materials, Singapore University of Technology and Design, Singapore 487372, Singapore}


\affiliation{Key Lab of Advanced Optoelectronic Quantum Architecture and Measurement (MOE), Beijing Key Lab of Nanophotonics $\&$ Ultrafine Optoelectronic Systems, and School of Physics, Beijing Institute of Technology, Beijing 100081, China}

\author{Xiaoting Zhou}

\affiliation{Department of Physics and Astronomy, California State University, Northridge, CA 91330, USA}

\author{Y. X. Zhao}

\address{National Laboratory of Solid State Microstructures and Department of Physics, Nanjing University, Nanjing 210093, China}

\address{Collaborative Innovation Center of Advanced Microstructures, Nanjing University, Nanjing 210093, China}

\author{Shengyuan A. Yang}

\address{Research Laboratory for Quantum Materials, Singapore University of Technology and Design, Singapore 487372, Singapore}

\begin{abstract}

Relativistic massless Weyl and Dirac fermions exhibit the isotropic and linear dispersion relations to preserve the Poincar\'{e} symmetry, the most fundamental symmetry in high energy physics. In solids, the counterparts of the Poincar\'{e} symmetry are crystallographic symmetries, and hence, it is natural to explore generalizations of Dirac and Weyl fermions compatible with their crystallographic symmetries and then the new physics coming along with them.
Here, we study an important kind of generalization, namely massless Dirac fermions with higher-order dispersion relations protected by crystallographic symmetries in three-dimensional nonmagnetic systems. We perform a systematic search over all 230 space groups with time-reversal symmetry and spin-orbit coupling considered. We find that the order of dispersion cannot be higher than three, i.e., only the quadratic and cubic Dirac points (QDPs and CDPs) are possible. We discover previously unknown classes of higher-order Dirac points, including the chiral QDPs with Chern numbers of $\pm 4$ and the QDPs/CDPs without centrosymmetry. Especially the chiral QDPs feature four extensive surface Fermi arcs and four chiral Landau bands and hence leads to observable signatures in spectroscopic and transport experiments. We further show that these higher-order Dirac points represent parent phases for other exotic topological structures. Via controlled symmetry breaking, QDPs and CDPs can be transformed into double Weyl points, triple Weyl points, charge-2 Dirac points or Weyl loops. Using first-principles calculations, we also identify possible material candidates, including $\alpha$-TeO$_2$ and YRu$_4$B$_4$, which realize the predicted nodal structures.
\end{abstract}
\maketitle

\section{Introduction}\label{Sec_intro}
Weyl and Dirac fermions are elementary particles in high energy physics, where the Poincar\'{e} symmetry is fundamental, and therefore massless Dirac fermions exhibit the isotropic twofold degenerate linear dispersion relation along all directions. Recently, topological metallic phases with protected
band degeneracies near the Fermi level have been attracting significant interest~\cite{Slager2013space-NP,Chiu2016Classification-RoMP,Burkov2016Topological-Nm,Bansil2016Colloquium-RoMP,Yan2017Topological-ARoCMP,Kruthoff2017Topological-PRX,Armitage2018Weyl-RoMP,Song2018Diagnosis-PRX,Zhang2019Catalogue-N,Vergniory2019complete-N,Tang2019Comprehensive-N,Tang2019Efficient-NP}, and particularly, seeking Weyl and Dirac fermions as quasiparticles in such condensed matter systems has been actively performed with various theoretical scenarios and material candidates being proposed~\cite{Murakami2007Phase-NJoP,Wan2011Topological-PRB,Young2012Dirac-Prl,Wang2012Dirac-PRB,Wang2013Three-PRB,Yang2016Dirac-spin,Shao2017Strain-PRB,Armitage2018Weyl-RoMP,Zhu2018Quadratic-PRB,Zhu2019Composite-PRB,Lu2019Robust-PRB}. Certainly, it is fascinating that the Lorentz-invariant massless Dirac (Weyl) fermions can be realized as quasiparticle excitation in the vicinity of fourfold (twofold) degenerate Fermi points with isotopic twofold degenerate (nondegenerate) linear dispersion, as various celebrated phenomena in high energy physics can lead to remarkable observable effects in Dirac (Weyl) semimetals~\cite{Nielsen1983Adler-PLB,Volovik2003universe-OUPoD,Guan2017Artificial-nQM}. However, in solids the most fundamental symmetry is the corresponding crystallographic symmetries, which are the counterpart of the Poincar\'{e} symmetry in high energy physics. Therefore, besides straightforwardly looking for faithful Dirac and Weyl quasiparticles, it is natural to properly extend the definitions of Dirac and Weyl fermions to be compatible with crystallographic symmetries. Since the linearity of dispersion for relativistic particles is required by the Lorentz symmetry, it should not be a surprise that in lots of cases crystallographic symmetries happen to contradict with the linear dispersion. In this respect, we focus in this paper on the Dirac points, which have higher order dispersions protected by crystallographic symmetries and are not limited to being twofold degenerate. Here, the term ``Dirac" only refers to the fourfold degeneracy of the Dirac point, following the previous convention~\cite{Yang2014Classification-Nc,Yang2015Topological-PRB,Gao2016Classification-PRB}.

With the generalization, we can go beyond previous schemes for Dirac semimetals to embrace more diversity for Dirac fermions and related novel physics. For instance, previous works are mostly focused on centrosymmetric nonmagnetic systems~\cite{Yang2014Classification-Nc,Yang2015Topological-PRB,Gao2016Classification-PRB,Liu2017Predicted-PRX,Yu2018Nonsymmorphic-PRM}, because it greatly simplifies the analysis: All the bands have an intrinsic Kramers degeneracy due to the combined $\mathcal{PT}$ symmetry [spin-orbital coupling (SOC) considered], such that a Dirac point is formed whenever two bands cross each other. However, this condition is not necessary for our generalized notion of Dirac fermions. As we shall see, for higher-order Dirac fermions, a variety of crystallographic symmetries can lead to fourfold degeneracy at some high-symmetry points in the Brillouin zone, typically with anisotropic behaviors. Especially, in the absence of $\mathcal{P}$, there are birefringent Dirac points with the four crossing bands fully splitting along certain directions, which were also proposed for linear Dirac fermions~\cite{Wang2013Three-PRB,Chen2017Ternary-PRM,Kennett2011Birefringent-PRA,Roy2012Asymmetric-PRB,Komeilizadeh2014Instabilities-PRB}.



Actually the importance of higher-order Dirac points has been noticed in a few scattered theoretical works. For example, Yang \emph{et al.} classified Dirac points on a rotational axis or at time reversal invariant momentum (TRIM) points for systems with both time reversal ($\mathcal{T}$) and inversion ($\mathcal{P}$) symmetries~\cite{Yang2014Classification-Nc}. Gao \emph{et al.} performed similar analysis and considered the constraints from seventeen different point groups~\cite{Gao2016Classification-PRB}. Both works reported the possible existence of special Dirac points with linear band splitting in one direction and
quadratic/cubic splitting in the orthogonal plane. Such Dirac points hence can be named as quadratic and cubic Dirac points (QDPs and CDPs). Notably, the CDPs were later predicted in realistic materials, such as Tl(MoTe)$_3$~\cite{Liu2017Predicted-PRX} and LiOsO$_3$~\cite{Yu2018Nonsymmorphic-PRM}.

Despite the progress mentioned above, higher-order Dirac points have not been thoroughly studied yet, and therefore
our current understanding of them is still limited. Particularly, it is worth emphasizing the importance of taking the whole crystallographic groups into consideration. A fundamental weakness in the previous works is that only certain subsets of the full crystallographic groups were considered, but the remaining symmetries may cause serious problems for the existence of Dirac fermions possibly from the following aspects. First, certain additional symmetries, although they do not affect the degeneracy at a Dirac point, may generate nodal lines or nodal surfaces that cover the fourfold degenerate point~\cite{Wu2018Nodal-PRB,Yu2019Circumventing-PRB}. Second, the nonsymmorphic symmetries, such as screw axis and glide mirror, have not been fully considered in previous studies, and their existence may strongly affect the symmetry conditions for Dirac points~\cite{Yu2018Nonsymmorphic-PRM}.



Motivated by these questions and challenges, here, we present a systematic investigation of higher-order Dirac points in three-dimensional (3D) systems. The time-reversal symmetry is assumed, and the SOC is fully considered. We search through all the 230 SGs of nonmagnetic materials, looking for higher-order Dirac points stabilized at the high-symmetry points of the Brillouin zone (BZ). The results are listed in Table~\ref{Tab1}. Our key findings include the following: (i)
We find that beyond the linear Dirac point, QDP and CDP are the only stable possibilities, namely, there are no symmetry-protected Dirac points with leading order dispersion (along any direction) higher than the third order. (ii) We discover a \emph{chiral} QDP which carries a topological charge (Chern number) of $\pm 4$ in SG~92 and 96. Such kind of chiral higher-order Dirac point has not been known before. (iii) We show that both QDP and CDP can be realized in crystals \emph{without} $\mathcal{P}$ symmetry, as in SG~92, 96, and 184-186. For these cases, the four bands generally split along generic directions deviating from the point. This behavior is in sharp contrast to the previously studied QDPs/CDPs.
(iv) For SG~142 and 228, although $\mathcal{P}$ is preserved, we find that QDP can be realized at points ($P$ and $W$, respectively) other than the TRIM points.

Furthermore, for each case in Table~\ref{Tab1}, we present the $k\cdot p$ effective model to characterize the low-energy emergent fermions. We discuss the physical signatures for the chiral QDP discovered here, including topological surface Fermi arcs and chiral Landau bands. {We further explore the possible topological phase transitions for QDPs and CDPs, and show that they may transform into double Weyl points, triple Weyl points, charge-2 Dirac points~\cite{Zhang2018Double-Prl}, or Weyl loops under proper symmetry breaking.} With first-principles calculations, we also identify concrete material examples including $\alpha$-TeO$_2$ and YRu$_4$B$_4$ for realizing some interesting cases in Table~\ref{Tab1}. Our work not only reveals previously unknown types of higher-order Dirac points, it also provides concrete symmetry guidelines for the materials search. Together with previous efforts~\cite{Yang2014Classification-Nc,Yang2015Topological-PRB,Gao2016Classification-PRB}, as far as we can see, this completes the classification of all possible higher-order Dirac points for 3D nonmagnetic systems. Some remaining questions and possible future directions are commented at the end.


{\renewcommand{\arraystretch}{1.8}

	\begin{table*}

		\caption{\label{Tab1} List of SGs hosting the quadratic and cubic Dirac points. The column with ``centrosymmetric" indicates whether the SG contains the centrosymmetry. The $\mathcal{C}$ in the penultimate column is the Chern number of the point.}

		\begin{ruledtabular}

			\begin{tabular}{ccccccccc}


				Order &  SG & BZ & Location  & Generators $\left\{ {\cal O}\big|t_{1}t_{2}t_{3}\right\} $ & Centrosymmetric  & $|\mathcal{C}|$ & Materials \tabularnewline

				\hline 

				\multirow{4}{*}{Quadratic} & 92 & \multirow{2}{*}{${\rm \Gamma_{q}}$} & \multirow{2}{*}{A $\left\{\frac{1}{2}\frac{1}{2}\frac{1}{2}\right\} $} & $\left\{ C_{4z}^{+}\big|00\frac{1}{4}\right\} $, $\left\{ C_{2;1\bar{1}0}\big|\frac{1}{2}\frac{1}{2}\frac{3}{4}\right\} $, $\mathcal{T}$ & \multirow{2}{*}{N} &\multirow{2}{*}{$4$} & $\alpha$-TeO$_{2}$ \tabularnewline


				& 96 &  &  & $\left\{ C_{4z}^{+}\big|00\frac{3}{4}\right\} $, $\left\{ C_{2;1\bar{1}0}\big|\frac{1}{2}\frac{1}{2}\frac{1}{4}\right\} $, $\mathcal{T}$ & & & \tabularnewline


				& 142

				& ${\rm \Gamma_{q}^{v}}$ & P $\left\{\frac{1}{4}\frac{1}{4}\frac{1}{4}\right\} $ & $\left\{ S_{4z}^{+}\big|\frac{1}{2}\frac{1}{2}\frac{1}{2}\right\} $, $\left\{ M_{1\bar{1}0}\big|10\frac{1}{2}\right\} $, $\mathcal{T}\left\{ \mathcal{P}|000\right\} $ & \multirow{2}{*}{Y} & \multirow{2}{*}{$0$} & YRu$_{4}$B$_{4}$ \tabularnewline


				& 228

				& ${\rm \Gamma_{c}^{f}}$ & W $\left\{\frac{1}{2}\frac{1}{4}\frac{3}{4}\right\} $ & $\left\{ S_{4x}^{+}\big|\frac{1}{2}\frac{1}{2}\frac{1}{2}\right\} $, $\left\{ M_{z}\big|\frac{3}{4}\frac{3}{4}\frac{3}{4}\right\} $, $\mathcal{T}\left\{\mathcal{P}|\frac{3}{4}\frac{3}{4}\frac{3}{4}\right\} $ & & & \tabularnewline

				\hline 

				\multirow{11}{*}{Cubic} &  184 & \multirow{3}{*}{${\rm \Gamma_{h}}$} & \multirow{3}{*}{A $\left\{00\frac{1}{2}\right\} $} & $\left\{ C_{6z}^{+}\big|000\right\} $, $\left\{ M_{y}\big|00\frac{1}{2}\right\} $, $\mathcal{T}$ & \multirow{3}{*}{N}  &\multirow{11}{*}{$0$} & \tabularnewline

				& 185 &  &  & $\left\{ C_{3z}^{+}\big|000\right\} $, $\left\{ C_{2z}\big|00\frac{1}{2}\right\} $, $\left\{ M_{y}\big|000\right\} $, $\mathcal{T}$ & & \tabularnewline


				& 186 &  &  & $\left\{ C_{3z}^{+}\big|000\right\} $, $\left\{ C_{2z}\big|00\frac{1}{2}\right\} $, $\left\{ M_{x}\big|000\right\} $, $\mathcal{T}$ & & \tabularnewline

				\cline{2-6}

				& 163 & \multirow{2}{*}{${\rm \Gamma_{h}}$} & \multirow{2}{*}{A $\left\{00\frac{1}{2}\right\} $} & $\left\{ C_{3z}^{+}\big|000\right\} $, $\left\{ C_{2y}\big|00\frac{1}{2}\right\} $, $\left\{\mathcal{P}|000\right\} $, $\mathcal{T}$ & \multirow{7}{*}{Y} & & \tabularnewline


				& 165 &  &  & $\left\{ C_{3z}^{+}\big|000\right\}$, $\left\{ C_{2x}\big|00\frac{1}{2}\right\} $, $\left\{\mathcal{P}|000\right\} $, $\mathcal{T}$ & & \tabularnewline


				& 167 & ${\rm \Gamma_{rh}}$ & Z $\left\{\frac{1}{2}\frac{1}{2}\frac{1}{2}\right\} $ & $\left\{ C_{3z}^{+}\big|000\right\} $, $\left\{ C_{2y}\big|\frac{1}{2}\frac{1}{2}\frac{1}{2}\right\} $, $\left\{\mathcal{P}|000\right\} $, $\mathcal{T}$ &  & & LiOsO$_{3}$\cite{Yu2018Nonsymmorphic-PRM} \tabularnewline


				& 226 & \multirow{2}{*}{${\rm \Gamma_{c}^{f}}$} & \multirow{2}{*}{L $\left\{\frac{1}{2}\frac{1}{2}\frac{1}{2}\right\} $} & $\left\{ C_{3,111}^{+}\big|000\right\} $, $\left\{ C_{2,1\bar{1}0}\big|000\right\} $, $\left\{ \mathcal{P}|\frac{1}{2}\frac{1}{2}\frac{1}{2}\right\} $, $\mathcal{T}$ &  \tabularnewline


				& 228 &  &  & $\left\{ C_{3,111}^{+}\big|000\right\} $, $\left\{ C_{2,1\bar{1}0}\big|\frac{1}{4}\frac{1}{4}\frac{1}{4}\right\} $, $\left\{\mathcal{P}|\frac{3}{4}\frac{3}{4}\frac{3}{4}\right\} $, $\mathcal{T}$ &   \tabularnewline


				& 176\footnote{The CDP in this SG is not an isolated point but resides on the intersection of three movable but unremovable nodal lines in the $k_{z}=\pi$ plane~\cite{ZhangPRM2018}.} & \multirow{2}{*}{${\rm \Gamma_{h}}$} & \multirow{2}{*}{A $\left\{00\frac{1}{2}\right\} $} & $\left\{ C_{6z}^{+}\big|00\frac{1}{2}\right\} $, $\left\{\mathcal{P}|00\frac{1}{2}\right\} $, $\mathcal{T}$ &  & & Tl(MoTe)$_{3}$\cite{Liu2017Predicted-PRX} \tabularnewline


				
				& 192 & & & $\left\{ C_{6z}^{+}\big|000\right\} $, $\left\{ M_{y}\big|00\frac{1}{2}\right\} $, $\left\{\mathcal{P}\big|000\right\} $, $\mathcal{T}$ &   \tabularnewline
				


			\end{tabular}

		\end{ruledtabular}

	\end{table*}

}

Before diving into detailed analysis, let us briefly comment on other promising directions beyond conventional Weyl and Dirac fermions. First, one can think of degenerate Fermi points forming higher dimensional manifolds, such as nodal lines (which can take the form of nodal rings~\cite{Yang2014Dirac-Prl,Weng2015Topological-PRB,Mullen2015Line-Prl,Yu2015Topological-Prl,Kim2015Dirac-Prl,Chen2015Nanostructured-Nl,Fang2016Topological-CPB,Li2017Type-PRB,Wu2019Hourglass-PRM,Shao2019Composite-NCM}, nodal links~\cite{Chang2017Topological-Prl}, nodal chains~\cite{Bzdusek2016Nodal-N,Wang2017Hourglass-Nc,Yu2017nodal-Prl}),
and even nodal surfaces~\cite{Zhong2016Towards-N,Liang2016Node-PRB,Wu2018Nodal-PRB,Zhang2018Nodal-PRB,Wang2018Pseudo-PRB,Gao2019Hexagonal-PRM,Fu2019Dirac-Sa,Yang2019Observation-NC}, especially with quadratic and cubic dispersion relations~\cite{Yu2019Quadratic-PRB}. Second, one can study different degrees of degeneracy, such as threefold and sixfold nodal points~\cite{Heikkilae2015Nexus-NJoP,Zhu2016Triple-PRX,Winkler2016Topological-Prl,Weng2016Coexistence-PRB,Weng2016Topological-PRB,Bradlyn2016Dirac-S,Hyart2016Momentum-PRB,Lv2017Observation-N,Chang2017Nexus-SR}. Third, another promising direction is to explore the interplay between crystallographic-symmetry representations and anisotropic dispersion relations, which would extend previous examples in the context of Weyl points. It has been shown that certain rotational symmetries can enforce the band splitting to be linear along the rotation axis, whereas in the plane orthogonal to the axis, the leading order splitting is of quadratic or cubic order~\cite{Xu2011Chern-Prl,Fang2012Multi-Prl,Tsirkin2017Composite-PRB}.

\section{approach} \label{Sec_approach}

Unlike the Weyl points which are topologically protected as long as the discrete translational symmetry is preserved, the Dirac points must require additional crystal symmetry protection. The condition is more stringent for higher-order Dirac points, because they require some symmetries to eliminate the lower-order terms in the band energy splitting. Therefore, such Dirac points have to reside at high symmetry points or on high-symmetry axis of the BZ. The case with high-symmetry axis has been discussed in previous works~\cite{Yang2014Classification-Nc,Yang2015Topological-PRB,Gao2016Classification-PRB}, so here we focus on the high-symmetry points. Distinct from the previous approaches, here we fully consider the SG symmetry, including the nonsymmorphic operations that play a crucial role for degeneracies at high-symmetry points on the BZ boundary~\cite{Young2012Dirac-Prl,Young2015Dirac-Prl,Wang2017Hourglass-Nc,Li2018Nonsymmorphic-PRB,Yu2018Nonsymmorphic-PRM}. As we have mentioned in the Introduction, this reveals new types of higher-order Dirac points which have not been reported before.

The search approach is similar to the one developed in our previous work~\cite{Yu2019Quadratic-PRB}. For each SG, we scan through its high-symmetry points, and look for symmetry-protected fourfold degeneracy. This is inferred from the dimension of the irreducible representations (IRRs) of the little group at the point. The generators of the little group can be found, e.g., in Ref.~\cite{Bradley2009Mathematical-Oxford}. The matrix representations of the symmetry operators can be established by analyzing the algebra formed by these operators.
Since we are concerned with nonmagnetic systems with SOC, we deal with the double-valued SG representations,
where a $2\pi$-rotation produces a minus sign and $\mathcal{T}^{2}=-1$.
Then, for each four dimensional IRR, we construct the most general symmetry-allowed ${k \cdot p}$ Hamiltonian expanded around the degeneracy point, from which the order of the Dirac point can be directly read off.
This procedure is applied to all the 230 SGs, which leads to the results presented in Table~\ref{Tab1}. The detailed derivations of the effective models for these higher-order Dirac points are presented in Supplemental Material (SM)~\cite{SupplementaryMaterial}. The obtained models are also double checked by using the kdotp-symmetry tool~\cite{Gresch2018Identifying-EZ}. In the following section, we shall use detailed examples to illustrate the approach.

\section{Quadratic Dirac point}\label{Sec_QDP}

\subsection{Chiral QDP}\label{SubSec_ChiralQDP}

As an illustration, let us consider SG~92, which can host a chiral QDP locating at the $A$ point. The little group at $A$ contains three generators: a screw rotation along the $z$ axis $\tilde{C}_{4z} \equiv \{C_{4z}^{+}|00\frac{1}{4}\} $ and a rotation axis along the $(1\bar{1}0)$ direction $C_{2;1\bar{1}0}\equiv\{ C_{2;1\bar{1}0}|\frac{1}{2}\frac{1}{2}\frac{3}{4}\}$, as well as $\mathcal{T}$. $\tilde{C}_{4z}$ and $C_{2;1\bar{1}0}$ satisfy the following algebra at $A$:
\begin{equation}\label{1}
\tilde{C}_{4z}^{4}=1, \ \  C_{2;1\bar{1}0}^{2}=-1, \ \ \tilde{C}_{4z}C_{2;1\bar{1}0}=-C_{2;1\bar{1}0}\tilde{C}_{4z}^{3}.
\end{equation}
The Bloch states at $A$ can be chosen as the eigenstates of $\tilde{C}_{4z}$, which we denote as $|c_{4z}\rangle$ with $c_{4z}\in\{\pm1,\ \pm i\}$ the eigenvalue of $\tilde{C}_{4z}$.
Based on Eq.~(\ref{1}), one finds that
\begin{equation}
\tilde{C}_{4z}C_{2;1\bar{1}0}|\pm1\rangle=-C_{2;1\bar{1}0}\tilde{C}_{4z}^{3}|\pm1\rangle=\mp C_{2;1\bar{1}0}|\pm1\rangle,
\end{equation}
which indicates that $C_{2;1\bar{1}0}|\pm1\rangle=|\mp1\rangle$ and the two states $|1\rangle$ and $|-1\rangle$ would be degenerate.
In addition, since ${\cal{T}}^2=-1$, the state $|\pm1\rangle$ and its time-reversal partner ${\cal{T}}|\pm1\rangle$ are linearly independent.
Hence, the four states $\{|1\rangle,\ C_{2;1\bar{1}0}|1\rangle,\ \mathcal{T}|1\rangle,\ \mathcal{T}C_{2;1\bar{1}0}|1\rangle\}$ must be degenerate at the same energy, forming a Dirac point.

To fully characterize this Dirac point and the associated emergent fermions, we construct the $k\cdot p$ effective model based on the symmetry.
The matrix representations of the generators can be expressed in the above quartet basis as
{
\begin{equation}
\tilde{C}_{4z} = \sigma_{0}\otimes\sigma_{z},\
C_{2;1\bar{1}0} = i\sigma_{0}\otimes\sigma_{y},\
\mathcal{T} = i\sigma_{y}\otimes\sigma_{0}\mathcal{K},
\end{equation}
where $\mathcal{K}$ is the complex conjugation, $\sigma_{i}$ ($i=x,y,z$) are the Pauli matrices, and $\sigma_{0}$ is the $2\times2$ identity matrix.} 
The Hamiltonian $\mathcal{H}_{\text{eff}}$ is required to be invariant under the symmetry transformations, namely,
\begin{eqnarray}
\tilde{C}_{4z}\mathcal{H}_{\text{eff}}(R_{4z}^{-1}\bm{k})\tilde{C}_{4z}^{-1} & = & {\mathcal{H}}_{\text{eff}}(\bm{k}),\label{eq:S4z}\\
C_{2;1\bar{1}0}{\mathcal{H}}_{\text{eff}}(R_{2;1\bar{1}0}^{-1}\bm{k})C_{2;1\bar{1}0}^{-1} & = & {\mathcal{H}}_{\text{eff}}(\bm{k}),\label{eq:S2b}\\
\mathcal{T}{\mathcal{H}}_{\text{eff}}(-\bm{k})\mathcal{T}^{-1} & = & {\mathcal{H}}_{\text{eff}}(\bm{k}),\label{eq:T}
\end{eqnarray}
where $\bm{k}$ is measured from the Dirac point, and $R_{4z}$ and $R_{2;1\bar{1}0}$ are the corresponding rotations acting on $\bm{k}$.
One notes that from Eq.~(\ref{eq:S4z}) together with $\tilde{C}_{4z}^{2}=\sigma_{0}\otimes\sigma_{0}$, the Hamiltonian must satisfy ${\mathcal{H}}_{\text{eff}}(-k_{x},-k_{y},k_{z})={\mathcal{H}}_{\text{eff}}(\bm{k})$. This clearly eliminates terms which are odd in $k_x$ and $k_y$,  indicating that the Dirac point might be a QDP.

It is convenient to write the $4\times 4$ model in the following block form
\begin{equation}\label{eq:model}
\mathcal{H}_{\text{eff}}^{\text{SG}}(\bm{k}) = w(\bm{k})\text{I}_{4\times 4} + \left[
\begin{array}{cc}
h_{11}^{\text{SG}}(\bm{k}) & h_{12}^{\text{SG}}(\bm{k}) \\
h_{12}^{\text{SG}\dagger}(\bm{k}) & h_{22}^{\text{SG}}(\bm{k})
\end{array}
\right]
\end{equation}
where each entry in the bracket is a $2 \times 2$ matrix. The $w$ term represents an overall energy shift for all the bands, which does not affect the order of the Dirac point. Hence, we will neglect the $w$ term in the following discussion.

With the constraint in Eqs.~(\ref{eq:S4z})-(\ref{eq:T}), the effective model expanded up to the second order is given as
{
\begin{eqnarray}
h_{11}^{92}(\bm{k}) & = & c_{1}k_{z}\sigma_{z}+[(c_{2}k_{+}^{2}+c_{3}k_{-}^{2})\sigma_{+} + \text{H.c.}],\label{eq:g92}\\
h_{12}^{92}(\bm{k}) & = & \alpha_{1}k_{z}\sigma_{z}+\alpha_{2}k_{x}k_{y}\sigma_{y},\label{eq:f92}
\end{eqnarray}
}
and
\begin{eqnarray}\label{10}
h_{22}^{92}(\bm{k})=h_{11}^{92*}(-\bm{k})
\end{eqnarray}
is a time-reversed partner of $h_{11}^{92}$. Here, $k_{\pm}=k_{x}\pm ik_{y}$ and $\sigma_{\pm}=(\sigma_{x}\pm i\sigma_{y})/2$. Note that here and hereafter, we use the Roman letters (such as $c_{i}$) and Greek letter (such as $\alpha_{i}$) to denote the real and complex parameters, respectively. This effective model confirms that the Dirac point is a QDP, with linear band splitting along $k_z$ and quadratic splitting in the $k_x$-$k_y$ plane.

More importantly, the diagonal blocks $h_{11}^{92}$ and its time-reversed partner $h_{22}^{92}$ each corresponds to a double Weyl point, and they share the \emph{same} topological charge (Chern number) of $2\rm{sgn}(|c_{2}|-|c_{3}|)$. Therefore, the Dirac point is chiral and has topological charge $\mathcal{C}=\pm 4$. To the best of our knowledge, such a chiral QDP has not been discovered before. Indeed, because the previous studies assume the inversion symmetry, the combined $\mathcal{PT}$ symmetry requires that the Berry curvature and hence the topological charge for any nodal point must be zero. In contrast, the QDP here can be chiral, because the SG considered here explicitly breaks the inversion symmetry. Chiral nodal points and the associated chiral emergent fermions are fascinating subject of research. 
For example, several types of unconventional chiral fermions have been proposed in chiral crystals~\cite{Chang2017Unconventional-Prl,Tang2017Multiple-Prl,Zhang2018Double-Prl,Chang2018Topological-Nm}. 
Later in Sec.~\ref{Sec_ChiralQDP_sig}, we shall discuss the interesting physics of this chiral QDP.

Similar analysis applies for SG~96, which results in a $k\cdot p$ model $\mathcal{H}_{\text{eff}}^{96}$ of the same form as SG~92 in Eqs.~(\ref{eq:g92})-(\ref{10}).
This proves that the Dirac point in SG~96 is also a chiral QDP with $\mathcal{C}=\pm 4$.

\subsection{QDP at non-TRIM point}\label{SubSec_QDP_nonTRIM}

Table~\ref{Tab1} shows that QDP may also be realized in SG~142 and 228. However, these two are distinct from SG~92 and 96, as the QDP here appears at a point ($P$ or $W$) that is not a TRIM point. As a result, the $\mathcal{T}$ symmetry does not belong to the little group at the location of the QDP, and hence it is \emph{not} a symmetry that protects the Dirac point. Nevertheless, as shown in Table~\ref{Tab1}, the combination of $\mathcal{T}$ with inversion does play an important role in stabilizing the Dirac point.

Following a similar method as in the last subsection, we find that the effective model for the QDP in SG~142 reads~\cite{SupplementaryMaterial}
\begin{eqnarray}
	h_{11}^{142}(\bm{k}) & = & e^{-i\pi/4}(ic_{1} k_{z} + c_{2}k_{+}^{2}+c_{3}k_{-}^{2})\sigma_{+} + \rm{H.c.},\\
	h_{12}^{142}(\bm{k}) & = & (\alpha_{1}k_{z}+\alpha_{2}k_{x}k_{y})\sigma_{y},
\end{eqnarray}
and
\begin{equation}
	h_{22}^{142}(\bm{k})=h_{11}^{142*}(\bm{k}).
\end{equation}
This QDP (and in fact all the bands for SG~142) is not chiral, because the SG contains both $\mathcal{T}$ and $\mathcal{P}$ symmetries.
As for the QDP in SG 228, we find that its effective model $\mathcal{H}_{\text{eff}}^{228}$ actually takes the same form as that for SG 142 (connected by a coordinate transformation and a unitary transformation~\cite{SupplementaryMaterial}).

Before proceeding, we comment that the symmetry $\mathcal{PT}$ which protects the QDP here is a kind of magnetic symmetry, meaning that it may also be preserved in a magnetic systems which explicitly breaks the $\mathcal{T}$ symmetry. Particularly, this $\mathcal{PT}$ may be present in certain antiferromagnetic systems. Previous works have shown that an antiferromagnet with $\mathcal{PT}$ symmetry may host a linear Dirac point~\cite{Tang2016Dirac-NP,Simejkal2017Electric-PRL,Wang2017Antiferromagnetic-PRB,Lin2019Dirac-apa}. Our discussion here suggests that it is also possible to realize a QDP in the presence of antiferromagnetic ordering with the required symmetry (as in Table~\ref{Tab1}). This would be an interesting topic for future studies.

\begin{figure*}
	\includegraphics[width=17cm]{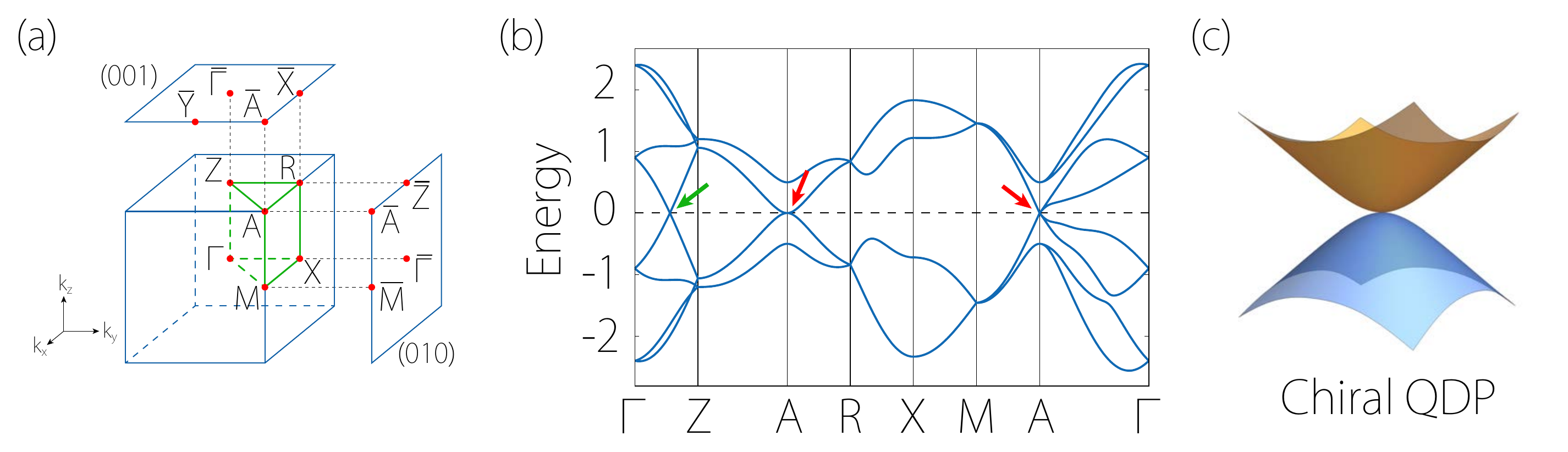}
	\caption{(a) Bulk and surface BZs for SG~92. (b) Bulk band structure for a lattice model with SG~92. The chiral QDP is located at the $A$ point (as indicated by the red arrows) and it carries a Chern number of $-4$. There also exists a pair of double Weyl points on the $k_z$ axis (as indicated by the green arrow), each with a Chern number of $+2$. The quadratic dispersion around the QDP in the plane perpendicular to $k_z$ in plotted in (c). \label{fig_f1}}
\end{figure*}

\section{Cubic Dirac point}\label{Sec_CDP}

The CDPs in Table~\ref{Tab1} can be put into two classes depending on whether the system contains the $\mathcal{P}$ symmetry or not. Below, we discuss the two classes one by one.

\subsection{CDP without inversion symmetry}\label{SubSec_CDP_noinv}

We first consider the CDP realized in SG~184-186, for which the inversion symmetry $\mathcal{P}$ is absent. These cases are distinct from the previously reported CDPs which all have the $\mathcal{P}$ symmetry. As a result, the four bands that form the CDP will fully split along a generic direction deviating from the point (twofold degeneracy may still appear along some high-symmetry direction such as the rotational axis).

For SG~184, the CDP resides at the $A$ point at the BZ boundary, and the nonsymmorphic symmetry such as $\{M_{y}|00\frac{1}{2}\}$ plays an important role in stabilizing the CDP. The corresponding effective Hamiltonian is obtained as
\begin{eqnarray}
h_{11}^{184}(\bm{k}) & = & f(\bm{k})k_{z}\sigma_{z}+[(\alpha_{1}k_{+}^{3}+\alpha_{2}k_{-}^{3})\sigma_{+}+{\rm H.c.}],\\
h_{12}^{184}(\bm{k}) & = & (c_{4}k_{+}^{3}+c_{5}k_{-}^{3})\sigma_{0}+[g(\bm{k})k_{z}\sigma_{+}+{\rm H.c.}],\\
h_{22}^{184}(\bm{k}) & = & h_{11}^{184}(k_x,-k_y,k_z),
\end{eqnarray}
where $f(\bm{k})=c_{1}+c_{2}k_{\parallel}^{2}+c_{3}k_{z}^{2}$, $g(\bm{k})=\alpha_{3}+\alpha_{4}k_{\parallel}^{2}+\alpha_{5}k_{z}^{2}$, and $k_{\parallel}=\sqrt{k_{x}^{2}+k_{y}^{2}}$.
One observes that the diagonal blocks $h_{11}^{184}(\bm{k})$ and $h_{22}^{184}(\bm{k})$ describe two triple Weyl points but with \emph{opposite}  Chern numbers $\pm3{\rm sgn}(|\alpha_{1}|-|\alpha_{2}|)$. Therefore,
the net Chern number of the CDP vanishes, which is consistent with the presence of the (glide) mirror that passing through this point.

For SG~185 and 186, the role of the glide mirror is replaced by a twofold screw rotation $\{C_{2z}|00\frac{1}{2}\}$. Together with the $\mathcal{T}$ symmetry, it actually ensures a twofold degeneracy in the $k_z=\pi$ plane.  As $A$ is located on this plane, a band crossing at $A$ will naturally have a fourfold degeneracy, making a Dirac point.
For SG~185, the effective Hamiltonian reads
\begin{eqnarray}
h_{11}^{185}(\bm{k}) & = & f(\bm{k})k_{z}\sigma_{0}+[i(c_{4}k_{+}^{3}+c_{5}k_{-}^{3})\sigma_{+} + \text{H.c.}],\\
h_{12}^{185}(\bm{k}) & = & g(\bm{k})k_{z}\sigma_{0} + \alpha_{4}(k_{+}^{3}-k_{-}^{3})\sigma_{x}, \\
h_{22}^{185}(\bm{k}) & = & h_{11}^{185*}(-\bm{k}).
\end{eqnarray}
where $f(\bm{k})=c_{1}+c_{2}k_{\parallel}^{2}+c_{3}k_{z}^{2}$, and $g(\bm{k})=\alpha_{1}+\alpha_{2}k_{\parallel}^{2}+\alpha_{3}k_{z}^{2}$.
Again, one finds that the diagonal blocks $h_{11}^{185}(\bm{k})$ and $h_{22}^{185}(\bm{k})$ describe two triple Weyl points with opposite Chern numbers $\pm3{\rm sgn}(|c_{4}|-|c_{5}|)$.

Similarly, the effective model of the CDP in SG~186 is related the model of SG~185 by a coordinate transformation, expressed as
\begin{eqnarray}
{\mathcal{H}}_{\text{eff}}^{186}(k_{x},k_{y},k_{z})&=&
{\mathcal{H}}_{\text{eff}}^{185}(-k_{y},k_{x},k_{z}).
\end{eqnarray}
Thus, although the inversion symmetry is broken for these SGs, the CDPs realized here are not chiral (i.e., with vanishing Chern numbers).

\subsection{CDP with inversion symmetry}\label{Sec_CDP_inv}

The remaining cases in Table~\ref{Tab1} are for the CDPs in SGs with inversion symmetry. These include SG~163, 165, 167, 226, 228, and 192. Due to the $\mathcal{PT}$ symmetry, the bands are at least doubly degenerate, and the Berry curvature as well as the net Chern number must vanish.
To construct the effective model, we note that for SG~163, 165, 167, 226, and 228, the inversion symmetry anticommute with the twofold rotation, and it satisfies $\mathcal{P}^{2}=1$. Hence, the fourfold degeneracy at the CDP can be decomposed into the four linearly independent states $\{|p=1\rangle,\ |p=-1\rangle,\ \mathcal{T}|p=1\rangle,\ \mathcal{T}|p=-1\rangle\}$, with $p$ denoting the eigenvalue of ${\cal{P}}$. Specifically, the effective model for SG~163 is obtained as
\begin{eqnarray}
h_{11}^{163}(\bm{k}) & = & [f(\bm{k})k_{z} + c_{4}k_{+}^{3} + c_{5}k_{-}^{3}]\sigma_{+} + \text{H.c.},\\
h_{12}^{163}(\bm{k}) & = & [g(\bm{k})k_{z}+\alpha_{4}(k_{+}^{3}+k_{-}^{3})]\sigma_{x}, \\
h_{22}^{163}(\bm{k}) & = & h_{11}^{163*}(-\bm{k}).
\end{eqnarray}
where $f(\bm{k})=c_{1}+c_{2}k_{\parallel}^{2}+c_{3}k_{z}^{2}$, and $g(\bm{k})=\alpha_{1}+\alpha_{2}k_{\parallel}^{2}+\alpha_{3}k_{z}^{2}$. Meanwhile, we find that the effective models for SG~165, 167, 226, and 228 all share the same form as SG~163 (after a coordinate transformation as expressed by Eqs.~(S76)-(S78) in SM~\cite{SupplementaryMaterial}).

At last, the effective model for the CDP in SG~192 is found to be
\begin{eqnarray}
h_{11}^{192}(\bm{k}) & = & (c_{1}k_{+}^{3} + c_{2}k_{-}^{3})\sigma_{+} + \text{H.c.},\\
h_{12}^{192}(\bm{k}) & = & g(\bm{k})k_{z}\sigma_{x}, \\
h_{22}^{192}(\bm{k})&=&h_{11}^{192*}(-\bm{k}).
\end{eqnarray}
where $g(\bm{k})=\alpha_{1}+\alpha_{2}(k_{x}^{2}+k_{y}^{2})+\alpha_{3}k_{z}^{2}$.

Previously, the CDP has been reported in two materials: One is in the centrosymmetric phase of LiOsO$_3$~\cite{Yu2018Nonsymmorphic-PRM}, the other is in the Tl(MoTe)$_3$ family materials~\cite{Liu2017Predicted-PRX}. The former case belongs to the SG~167, while the latter case belongs to the SG~176. It should be noted that different from the other cases, the CDP in SG~176 is not an isolated Dirac point. Instead, it resides on the intersection of three movable but unremovable Dirac nodal lines on the $k_z=\pi$ plane~\cite{ZhangPRM2018}. Therefore, strictly speaking, this degeneracy should not be classified as a Dirac point.

\section{signatures of chiral QDP} \label{Sec_ChiralQDP_sig}

We have demonstrated that different kinds of QDPs and CDPs can be stabilized by the space group symmetry. The most interesting discovery here is the chiral QDP which carries a nonzero Chern number $\pm 4$. Below, we focus on this case and discuss its two interesting physical signatures, including the topological surface Fermi arcs and the chiral Landau bands.

\subsection{Surface Fermi arcs}\label{SubSec_ChiralQDP_arcs}

\begin{figure}
	\includegraphics[width=8.5cm]{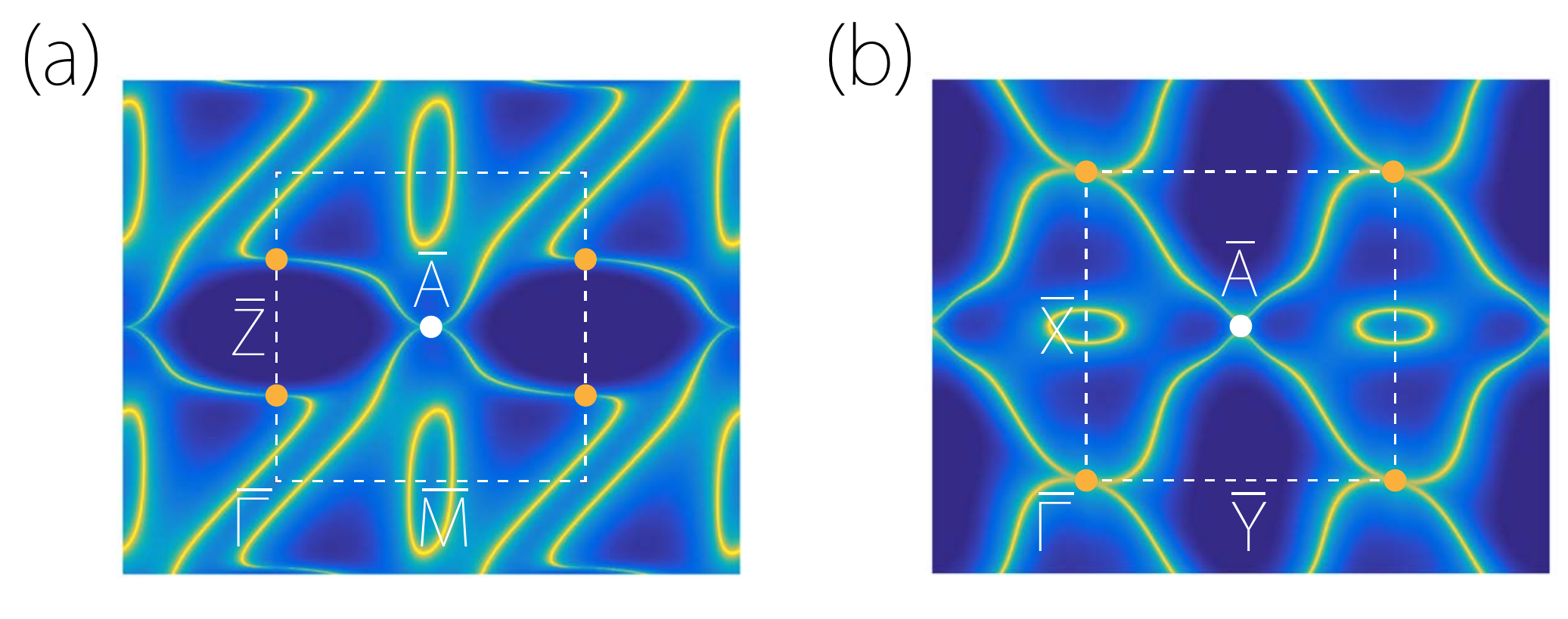}
	\caption{The Fermi surface contours on (a) the (010) surface and (b) the (001) surface for the lattice model with SG~92. The white and orange dots indicate the surface projections of the chiral QDP and the double Weyl points, respectively. There are four surface Fermi arcs emanating from the projection of the chiral QDP, consistent with its topological charge. \label{fig_f2}}
\end{figure}

Because the QDPs in SG~92 and 96 carry nonzero Chern number with absolute value $|\mathcal{C}|=4$, according to the bulk-boundary correspondence~\cite{Wan2011Topological-PRB}, on the surface of the material, there should exist four Fermi arcs emerging from the surface projection of the Dirac point. To explicitly demonstrate this,
we construct an eightband tight-binding model for SG~92 (see SM~\cite{SupplementaryMaterial}),
and calculate its bulk and surface spectra. In Fig.~\ref{fig_f1}(b), one observes the QDP located at the $A$ point. Figure~\ref{fig_f1}(c) shows the quadratic band dispersion around this QDP in the $k_{x}$-$k_{y}$ plane (here, each band is doubly degenerate because of the $\mathcal{T}\tilde{C}_{2z}$ symmetry). Figure~\ref{fig_f2}(a) shows the surface spectrum for the (010) surface, in which one can clearly observe four surface Fermi arcs emanating from the projected QDP. These arcs are terminated at the projections of two double Weyl points located on the $k_z$ axis [see Fig.~\ref{fig_f2}(a)].
For this tight-binding model, we find that the chiral QDP carries a Chern number of $-4$, and each double Weyl point carries a Chern number of $+2$, so the net topological charge in the BZ vanishes, satisfying the no-go theorem. It is also worth pointing out that because these chiral nodal points are sitting at different locations in the BZ ($A$ point and $k_z$ axis) which are far apart, the Fermi arcs connecting them are extensive in the surface BZ. This makes them more accessible in angle-resolved photoemission spectroscopy (ARPES) or scanning tunneling spectroscopy (STS) experiment.

\subsection{Chiral Landau bands}\label{SubSec_ChiralQDP_Landau}

\begin{figure}[t]
\includegraphics[width=8cm]{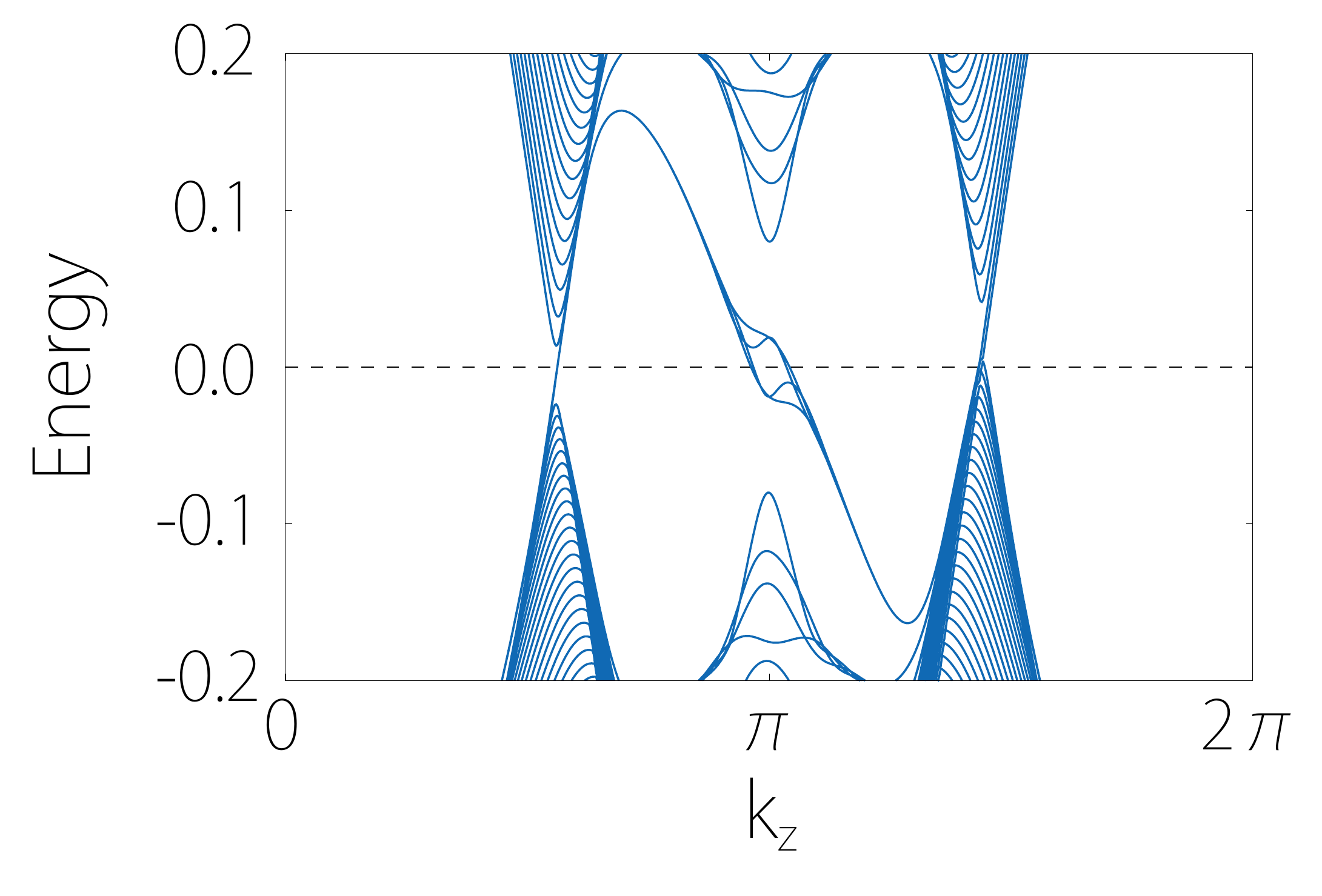}
\caption{Landau spectrum calculated from the lattice model for SG~92, with $B$ field along the $z$ direction. The chiral QDP gives four chiral Landau bands crossing the Fermi level with negative slope (near $k_z=\pi$ here). They are accomplished by the four positive chiral Landau bands associated with the two double Weyl points. \label{fig_f3}}
\end{figure}

The chiral nature of a nodal point can manifest in the spectrum under a strong magnetic field. For the conventional Weyl point with topological charge of $\pm 1$, there exists one chiral Landau band with linear dispersion along the magnetic field direction~\cite{Nielsen1983Adler-PLB}. In Ref.~\cite{Zhao2017Topological-apa}, it has been shown that the net number of chiral Landau bands just corresponds to the topological charge.
Here, since the QDP has a topological charge of $\pm 4$, one expects to find four such Landau bands.

In Fig.~\ref{fig_f3}, we show the explicit result of the Landau spectrum based on the tight-binding model. (An analytic solution of the $k\cdot p$ model is presented in SM~\cite{SupplementaryMaterial}.) Here, we take the same model as in Fig.~\ref{fig_f1} for SG~92, and the magnetic field is oriented along the $z$-axis. Under the magnetic field, the electron motion in the $x$-$y$ plane is quantized into Landau levels, so the original 3D band structure transforms into 1D Landau bands with dispersion only along $k_z$. In the spectrum, one clearly observes four chiral Landau bands with negative slopes around $k_z=\pi$, corresponding to the chiral QDP. Meanwhile, the two double Weyl points residing on the $k_z$ axis give another four chiral Landau bands with positive slopes. Similar to the Weyl case, when further applying an electric field parallel to the magnetic field, electrons will be pumped between the different chiral points, and this can result in a negative contribution to the longitudinal magnetoresistance~\cite{Son2013Chiral-PRB}. Recent works have shown that this chiral anomaly related process would undergo a breakdown when the magnetic field is strong enough such that the inverse magnetic length $\ell_{B}^{-1}$ is comparable to the chiral point separation~\cite{Ramshaw2018Quantum-Nc, Chan2017Emergence-PRB, Kim2017Breakdown-PRL}. For SG~92 and 96, since the QDP is well separated from the other nodal point in momentum space, one can expect that this chiral anomaly effect will be more robust against the magnetic tunneling.

%
	
{\renewcommand{\arraystretch}{1.8}
\begin{table*}
	{
	\caption{\label{Tab2}
		Topological phase transitions for higher-order Dirac points under symmetry breaking. WPs and DPs stand for the Weyl points and the Dirac points, respectively.}
	\begin{ruledtabular}
	\begin{tabular}{ccllc}
		Type & SG & Change of SG & Symmetry breaking & New phase\tabularnewline
		\hline
		\multirow{4}{*}{Chiral QDP} & 92 & $\Rightarrow$ 76 & $\{C_{2;1\bar{1}0}|\frac{1}{2}\frac{1}{2}\frac{3}{4}\}$ (retain $\{C_{4z}^{+}|00\frac{1}{4}\}$) & \multirow{2}{*}{A pair of double WPs at $(\pi,\pi,\pi\pm q_{z})$}\tabularnewline
		& 96 & $\Rightarrow$ 78 & $\{C_{2;1\bar{1}0}|\frac{1}{2}\frac{1}{2}\frac{1}{4}\}$ (retain $\{C_{4z}^{+}|00\frac{3}{4}\}$) & \tabularnewline
		\cline{2-5} \cline{3-5} \cline{4-5} \cline{5-5}
		& 92 & $\Rightarrow$ 19 & $\{C_{4z}^{+}|00\frac{1}{4}\}$ (retain $\{C_{2z}|00\frac{1}{2}\}$) & \multirow{2}{*}{A pair of charge-2 DPs}\tabularnewline
		& 96 & $\Rightarrow$ 19 & $\{C_{4z}^{+}|00\frac{3}{4}\}$ (retain $\{C_{2z}|00\frac{1}{2}\}$) & \tabularnewline
		\hline
		\multirow{2}{*}{non-TRIM QDP} & \multirow{2}{*}{142}  & $\Rightarrow$ 110 & $\mathcal{P}$ (retain $\{C_{2z}^{+}|\frac{1}{2}0\frac{1}{2}\}$ and $\{M_{1\bar{1}0}|10\frac{1}{2}\}$) & \multirow{2}{*}{Weyl chain} \tabularnewline
		& & $\Rightarrow$ 122 & $\mathcal{P}$ (retain $\{S_{4z}^{+}|\frac{1}{2}\frac{1}{2}\frac{1}{2}\}$ and $\{M_{1\bar{1}0}|10\frac{1}{2}\}$) & \tabularnewline
		\cline{2-5} \cline{3-5} \cline{4-5} \cline{5-5}
		\hline
		\multirow{5}{*}{non-$\mathcal{P}$ CDP} & 185 & $\Rightarrow$ 173 & $\{M_{y}|000\}$ & \multirow{2}{*}{A pair of triple WPs at $(0,0,\pi\pm q_{z})$}\tabularnewline
		& 186 & $\Rightarrow$ 173 & $\{M_{x}|000\}$ & \tabularnewline
		\cline{2-5} \cline{3-5} \cline{4-5} \cline{5-5}
		& 184 & $\Rightarrow$ 158/159 & $\{C_{2z}|000\}$ & \multirow{3}{*}{Weyl loops}\tabularnewline
		& 185 & $\Rightarrow$ 158 & $\{C_{2z}|00\frac{1}{2}\}$ (retain $\{M_{x}|00\frac{1}{2}\}$) & \tabularnewline
		& 186 & $\Rightarrow$ 159 & $\{C_{2z}|00\frac{1}{2}\}$ (retain $\{M_{y}|00\frac{1}{2}\}$) & \tabularnewline
		\hline
		\multirow{5}{*}{$\mathcal{P}$ CDP} & 163 & $\Rightarrow$ 159 & $\mathcal{P}$ (retain $\{M_{y}|00\frac{1}{2}\}$) & \multirow{5}{*}{Weyl loops}\tabularnewline
		& 165 & $\Rightarrow$ 158 & $\mathcal{P}$ (retain $\{M_{x}|00\frac{1}{2}\}$) & \tabularnewline
		& 167 & $\Rightarrow$ 161 & $\mathcal{P}$ (retain $\{M_{y}|\frac{1}{2}\frac{1}{2}\frac{1}{2}\}$) & \tabularnewline
		& 226/228 & $\Rightarrow$ 219 & $\mathcal{P}$ (retain $\{M_{1\bar{1}0}|\frac{1}{2}\frac{1}{2}\frac{1}{2}\}$) & \tabularnewline
		& 192 & $\Rightarrow$ 188/190 & $\mathcal{P}$ (retain $\{M_{x}|00\frac{1}{2}\}$/$\{M_{y}|00\frac{1}{2}\}$) & \tabularnewline
	\end{tabular}
    \end{ruledtabular}
}
\end{table*}
}
\begin{figure}[t]
	\includegraphics[width=8.5cm]{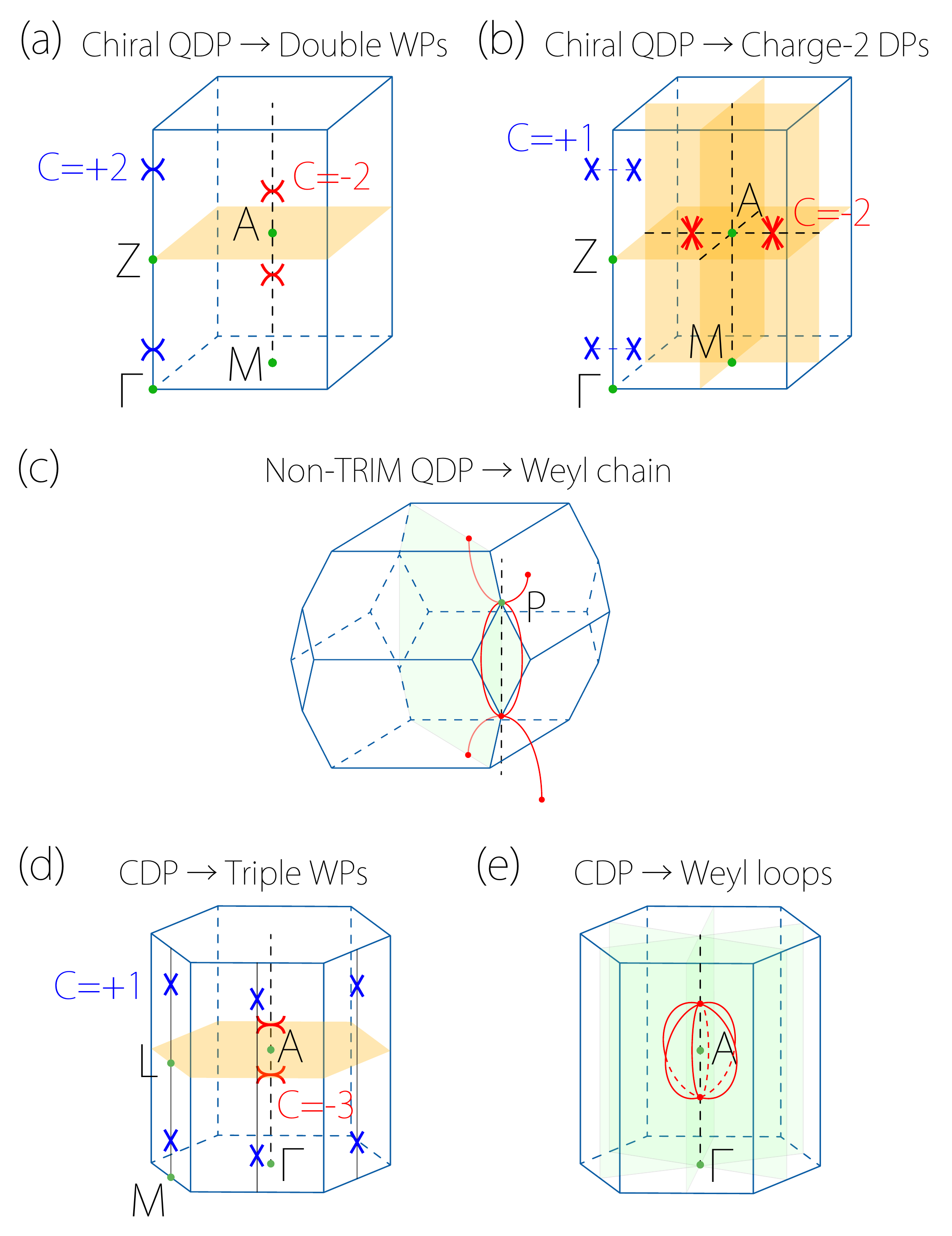}
	\caption{QDPs and CDPs under symmetry breaking. (a) Breaking $C_{2;1\bar{1}0}$ rotation axis can transform the chiral QDPs of SGs 92 and 96 into a pair of double Weyl points (red crosses), while the double Weyl points on the $k_z$ axis (blue crosses) remain. (b) By breaking the $\tilde{C}_{4z}$ screw axis (retaining $\tilde{C}_{2z}$), the chiral QDPs can be transformed into a pair of charge-2 Dirac points (red crosses), while the double Weyl points on the $k_z$ axis split into a pair of linear Weyl points (blue crosses). {(c) Breaking $\mathcal{P}$ (retaining $\{S_{4z}^{+}|\frac{1}{2}\frac{1}{2}\frac{1}{2}\}$ and $\{M_{1\bar{1}0}|10\frac{1}{2}\}$) can transform the QDP of SG 142 to nodal chains with the touching point located at the point P.} (d) Breaking $M_{y}$ mirror can transform the CDP of SG 185 into a pair of triple Weyl points (red crosses). Their topological charges are compensated by three pairs of linear Weyl points on the three $M$-$L$ path (blue crosses). (e) Breaking $\mathcal{P}$ (retaining $\tilde{M}_{x}$) can transform the CDP of SG 185 into three crossed Weyl loops lying in the three glide invariant planes.}
	\label{fig_tpt}
\end{figure}

\section{Topological phase transition}

As the higher-order Dirac points are protected by symmetry, they will generally be gapped or transformed to other types of band degeneracies when the symmetry is broken. Several interesting cases are illustrated in Table~\ref{Tab2} and Fig.~\ref{fig_tpt}.
For example, breaking the $C_{2;1\bar{1}0}$ rotation axis can split the chiral QDP of SGs 92 and 96 into a pair of double Weyl points with the same chirality [Fig.~\ref{fig_tpt}(a)]. For SGs 185 and 186, the breaking of mirror symmetry ($M_{y}$ or $M_{x}$) may transform the CDP into a pair of triple Weyl points [Fig.~\ref{fig_tpt}(d)]. The triple Weyl points have the same chirality, which would be compensated by three pairs of linear Weyl points with the opposite chirality on the three $M$-$L$ paths. For SGs 92 and 96, one can also obtain a pair of charge-2 Dirac points~\cite{Zhang2018Double-Prl} with the same Chern number $|\mathcal{C}|=2$ on a screw invariant axis, via breaking the $\tilde{C}_{4z}$ screw axis while retaining the $\tilde{C}_{2z}$ screw axis [Fig.~\ref{fig_tpt}(b)]. For SG 142, retaining $\{S_{4z}^{+}|\frac{1}{2}\frac{1}{2}\frac{1}{2}\}$ and $\{M_{1\bar{1}0}|10\frac{1}{2}\}$ while breaking $\mathcal{P}$, the QDP at non-TRIM can be transformed into nodal chains whose touching point is located at point P [Fig.~\ref{fig_tpt}(c)]. This is consistent with the previous study~\cite{Bzdusek2016Nodal-N}. Moreover, retaining a glide plane while breaking $C_{2z}$ or $\mathcal{P}$ that anticommute with the glide mirror can transform the CDP into Weyl loops traced by the necking point of hourglass dispersions. This explains the appearance of mutually crossed nodal rings in the ferroelectric phase of LiOsO$_{3}$~\cite{Yu2018Nonsymmorphic-PRM}.
These results demonstrate that QDP and CDP systems provide a promising playground for studying topological phase transitions and a variety of emergent fermions.

It should be mentioned that the list in Table~\ref{Tab2} is not exhaustive. There may exist other topological gapless or gapped phases realized by symmetry breaking from these higher-order Dirac points~\cite{Yang2014Classification-Nc,Wieder2016Double-Prl}. The transformation of higher-order Dirac points is an interesting topic which deserves further studies.

\section{Material realization} \label{Sec_Materials}

The symmetry conditions summarized in Table~\ref{Tab1} provide useful guides for the material search. As indicated in the table, the previously identified materials LiOsO$_3$~\cite{Yu2018Nonsymmorphic-PRM} and Tl(MoTe)$_3$~\cite{Liu2017Predicted-PRX} belong to two space groups which can host CDPs in the presence of $\mathcal{P}$ symmetry. Here, we identify two material examples that host the new kinds of higher-order Dirac points discovered in this work.

\begin{figure}[t]
\includegraphics[width=8.5cm]{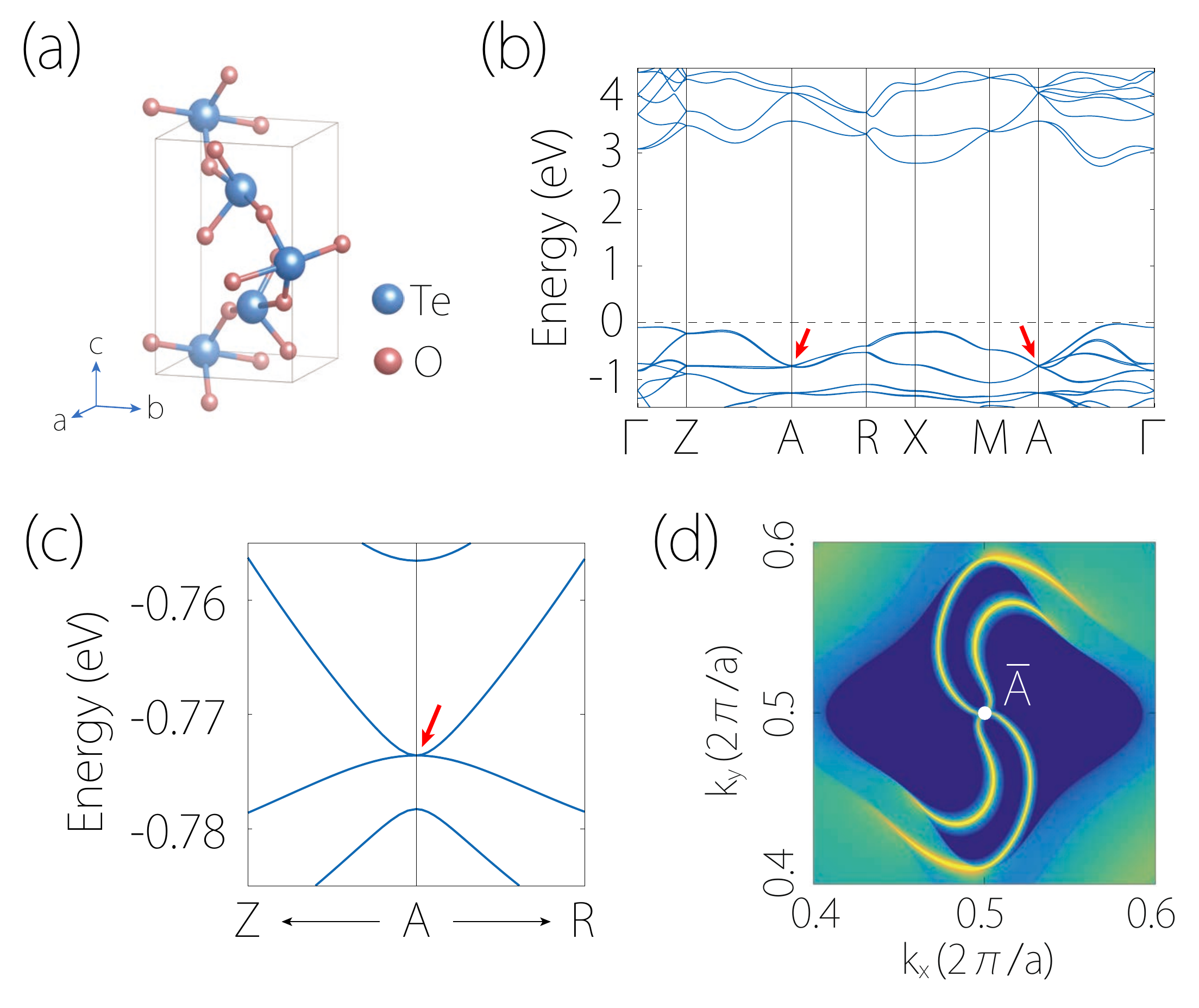}
\caption{(a) Crystal structure for $\alpha$-TeO$_{2}$ (SG~92). (b) Calculated band structure for $\alpha$-TeO$_{2}$ (with SOC included). The red arrows indicate the chiral QDP at $A$. (c) shows the zoom-in image around the QDP. (d) Calculated constant energy slice (at the QDP energy) for the (001) surface shows four surface Fermi arcs emitted from the projected chiral QDP.}
\label{fig_f4}
\end{figure}

The first example is the tetragonal paratellurite, $\alpha$-TeO$_{2}$, which belongs to SG~92. This material has been synthesized in experiment and exists as the low-pressure phase of TeO$_{2}$~\cite{Leciejewicz1961Crystal-ZfKCM,Lindqvist1968Refinement-ACS,Worlton1975Structure-PRB,Liu1987Polymorphism-JPCS,Thomas1988Crystal-JPCSSP}. Here, we only focus on the tetragonal low-pressure phase. Its structure is built up of asymmetric TeO$_{4}$ trigonal bipyramid polyhedron units [see Fig.~\ref{fig_f4}(a)]. We use first-principle calculations to obtain its band structure (SOC included), which has been plotted in Fig.~\ref{fig_f4}(b). Although it is an indirect band-gap semiconductor with a band gap of $\sim 2.69$~eV, the feature of QDPs can be found at the $A$ point in both conduction and valence bands. From a zoom-in image in Fig.~\ref{fig_f4}(c), one clearly observes the quadratic dispersion within the $k_{x}$-$k_{y}$ plane. The calculated Chern number for this QDP is $-4$, which is consistent with our symmetry analysis.
On a generic surface, there should be four Fermi arcs emerging from the surface projection of the QDP point. We calculate the projected spectrum for the (001) surface, and indeed verify the existence of four topological surface Fermi arcs, as shown in Fig.~\ref{fig_f4}(d).

\begin{figure}
	\includegraphics[width=8.5cm]{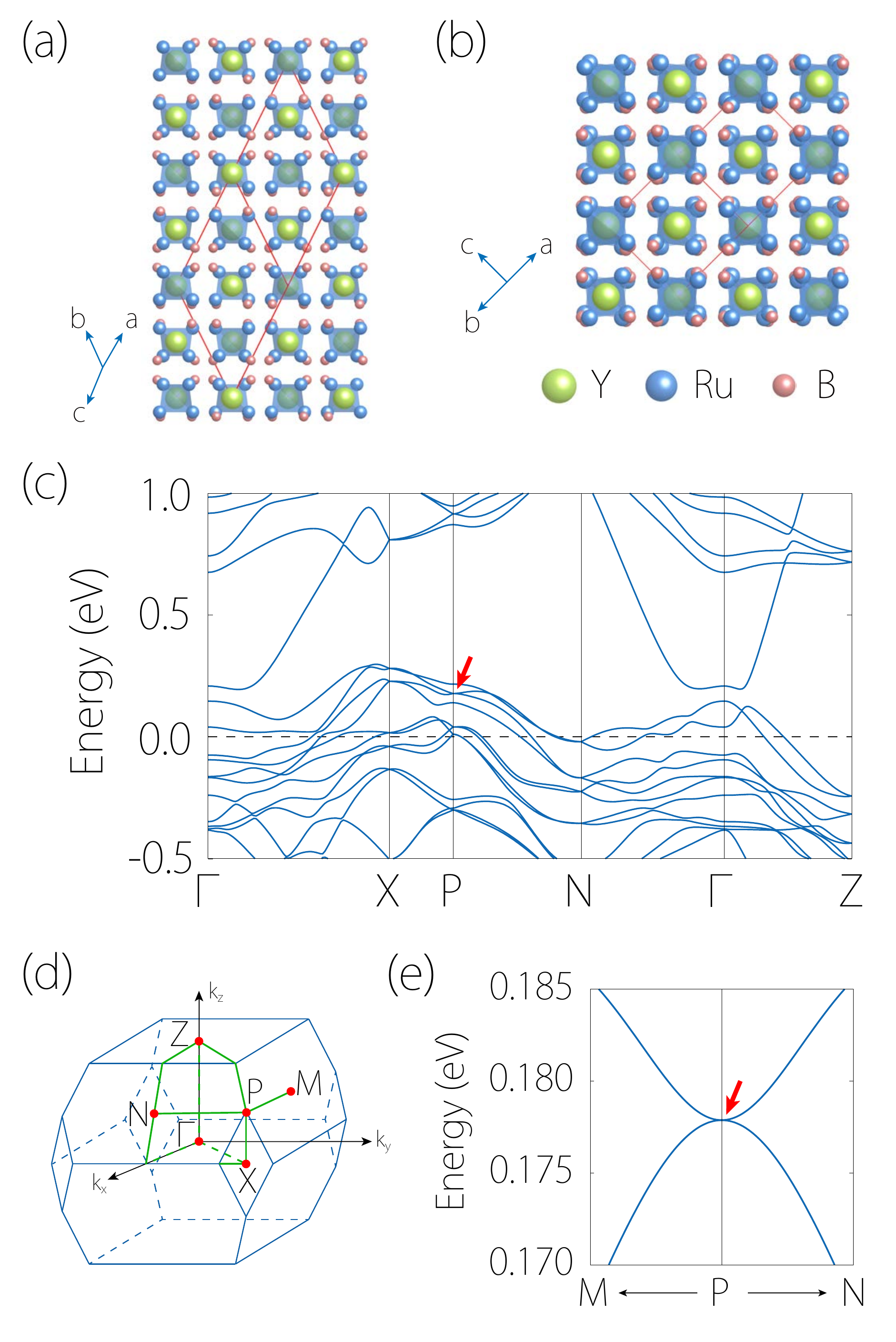}
	\caption{(a) Side and (b) top view of the crystal structure of YRu$_{4}$B$_{4}$. The red parallelograms indicate the primitive cell. (c) Calculated band structure for YRu$_{4}$B$_{4}$ (with SOC included). The red arrows indicate the QDP at $P$. (d) shows the BZ.
    (e) Enlarged view of the band dispersion around the QDP in (c).  }
	\label{fig_f5}
\end{figure}

The second example is YRu$_{4}$B$_{4}$, which is a member of the superconducting materials family $M$(Rh,Ru)$_{4}$B$_{4}$ ($M =$ Y, Th, or Lanthanides)~\cite{Johnston1977Superconductivity-SSC,Shelton1983Charge-PRB}. The material has a tetragonal crystal structure with the space group of $I4_{1}/acd$ (No.~142).
In the lattice structure, as shown in Fig.~\ref{fig_f5}(a) and~\ref{fig_f5}(b), the Y atoms form a face-centred-cubic sublattice, while Ru atoms form an array of tetrahedra with B atoms interspersed among them. According to Table~\ref{Tab1}, this SG can host a QDP at a non-TRIM point.
In the band structure of YRu$_{4}$B$_{4}$ plotted in Fig.~\ref{fig_f5}(c), one indeed observes a Dirac point about 0.18~eV above the Fermi level at the non-TRIM point $P$. The dispersion around the Dirac point [see Fig.~\ref{fig_f5}(e)] confirms that it has a quadratic in-plane dispersion, which is in agreement with our symmetry analysis.

\section{Discussion and Conclusion} \label{Sec_Discussion}

In this work, we have systematically investigated the higher-order Dirac points that can be stabilized at high-symmetry points for nonmagnetic systems. The fourfold degeneracy of a Dirac point must require protection from certain crystalline symmetry, so the Dirac point generally cannot appear at a generic $k$ point in the BZ; instead it might be located at a high-symmetry point (studied in this work), on a high-symmetry path (Ref.~\cite{Yang2014Classification-Nc,Yang2015Topological-PRB,Gao2016Classification-PRB}), or on a high-symmetry (mirror) plane. Regarding the last case, the previous work by Yang \emph{et al.}~\cite{Yang2014Dirac-Prl} have indicated that a combination of chiral symmetry, mirror symmetry, $\mathcal{P}$, and $\mathcal{T}$ may stabilize linear Dirac points on a mirror plane. The chiral symmetry is a natural symmetry for the superconducting quasiparticle spectrum, and it may also emerge at low energy for many elemental materials with a bipartite lattice~\cite{Chen2015Nanostructured-Nl,Zhong2016Towards-N,Lu2016Multiple-nCM,Zhong2017Three-Nc,Wu2018Nodal-PRB}. However, to ensure a higher-order dispersion, the condition is more stringent. Additional symmetries must be needed to rule out the linear order terms, and hence a higher-order Dirac point on a mirror plane appears unlikely. Therefore, we speculate that our work, together with previous studies~\cite{Yang2014Classification-Nc,Yang2015Topological-PRB,Gao2016Classification-PRB}, has examined all possible higher-order Dirac points in 3D nonmagnetic systems.

We have identified two material examples which contain the higher-order Dirac points. The purpose is to confirm our symmetry analysis. However, these materials are not ideal, in the sense that either the Dirac point is not close to the Fermi level (for $\alpha$-TeO$_2$), or the low-energy bands are not clean (for YRu$_{4}$B$_{4}$). To facilitate experimental research on these new nodal points, it is important to search out better material candidates in future studies. Our results here will provide a useful guidance for this task.

In conclusion, via symmetry analysis, we have searched through all 230 SGs and obtained all possible higher-order Dirac points at high-symmetry points for 3D nonmagnetic systems. We show that only QDP and CDP are possible, i.e., there is no stable Dirac point with dispersion higher than the third order. We find several types of previously unknown Dirac points, including the chiral QDP (with Chern number $\mathcal{C}=\pm 4$), QDP at non-TRIM point, and CDP without the $\mathcal{P}$ symmetry. We present the list of SGs that can host these Dirac points as well as their low-energy effective models. For the chiral QDP, we further discuss its interesting physical properties, including the extensive surface Fermi arcs and chiral Landau bands. {We also explore the topological phase transitions for QDPs and CDPs under symmetry breaking, and show that they can give rise to a rich variety of topological band degeneracies, such as double Weyl points, triple Weyl points, charge-2 Dirac points or Weyl loops.} Finally, we identify material examples $\alpha$-TeO$_2$ and YRu$_{4}$B$_{4}$ that exhibit the higher-order Dirac points. Our work discovers topological gapless phases with new kinds of emergent Dirac fermions. The obtained symmetry conditions will be useful to guide material search. The approach adopted here may also be extended to study new kinds of nodal structures in magnetic systems in future works.

\begin{acknowledgments}
The authors thank D. L. Deng for valuable discussions. This work is supported by the Singapore Ministry of Education AcRF Tier 2 (MOE2017-T2-2-108), the Fundamental Research Funds for the Central Universities (Grant No. 14380119), National Natural Science Foundation of China (Grant No. 11874201), and Beijing Institute of Technology Research Fund Program for Young Scholars. We acknowledge computational support from the Texas Advanced Computing Center and the National Supercomputing Centre Singapore.
\end{acknowledgments}


%

\end{document}


\title{Supplementary Materials for ``Higher-order Dirac fermions in three dimensions''}

\author{Weikang Wu}


\address{Research Laboratory for Quantum Materials, Singapore University of Technology and Design, Singapore 487372, Singapore}

\author{Zhi-Ming Yu}


\address{Research Laboratory for Quantum Materials, Singapore University of Technology and Design, Singapore 487372, Singapore}


\affiliation{Key Lab of Advanced Optoelectronic Quantum Architecture and Measurement (MOE), Beijing Key Lab of Nanophotonics $\&$ Ultrafine Optoelectronic Systems, and School of Physics, Beijing Institute of Technology, Beijing 100081, China}

\author{Xiaoting Zhou}

\affiliation{Department of Physics and Astronomy, California State University, Northridge, CA 91330, USA}

\author{Y. X. Zhao}

\address{National Laboratory of Solid State Microstructures and Department of Physics, Nanjing University, Nanjing 210093, China}

\address{Collaborative Innovation Center of Advanced Microstructures, Nanjing University, Nanjing 210093, China}

\author{Shengyuan A. Yang}

\address{Research Laboratory for Quantum Materials, Singapore University of Technology and Design, Singapore 487372, Singapore}

\maketitle

\section{First-principles calculation method}\label{ASec_method}

The calculation for real materials in this work was performed by the first-principle methods based on the density functional theory (DFT), as implemented in the Vienna \emph{ab initio} simulation package~\cite{Kresse1993Ab-PRB,Kresse1996Efficient-PRB}.  The projector augmented wave method was adopted~\cite{Bloechl1994Projector-PRB}. The generalized gradient approximation (GGA) with the Perdew-Burke-Ernzerhof (PBE)~\cite{Perdew1996Generalized-PRL} realization was adopted for the exchange-correlation potential. For all calculations, the energy and force convergence criteria were set to be $10^{-8}$~eV and $10^{-3}$~eV/\text{\AA}, respectively. The BZ sampling was performed by using $k$ grids with a spacing of $2\pi \times 0.02$ \AA$^{-1}$ within a $\Gamma$-centered sampling scheme. For $\alpha$-TeO$_{2}$, the optimized lattice constants are $a=4.971\text{\AA}$ and $c=7.620\text{\AA}$, which are close to the experimental values~\cite{Leciejewicz1961Crystal-ZfKCM,Lindqvist1968Refinement-ACS,Worlton1975Structure-PRB,Liu1987Polymorphism-JPCS,Thomas1988Crystal-JPCSSP}. For YRu$_{4}$B$_{4}$, the experimental values of lattice parameters were used in the calculation~\cite{Johnston1977Superconductivity-SSC,Shelton1983Charge-PRB}.
The surface states were investigated by constructing the maximally localized Wannier functions~\cite{Marzari1997Maximally-PrB,Souza2001Maximally-PRB,Mostofi2014updated-CPC} using the WANNIER-TOOLS package combined with an iterative Green's function method~\cite{Sancho1984Quick-JoPFMP,Sancho1985Highly-JoPFMP,Wu2018WannierTools-CPC}.

\section{Derivation of effective Hamiltonian}\label{ASec_kpmodel}
Here, we present the derivation of the effective models in more detail. For convenience, the $4\times 4$ models are written in the block form of Eq.~(7) in the main text.

Before derivation, we first introduce the notations to be used. A Bravais lattice is defined by three basis vectors, $\bm{t}_{i}$, and the reciprocal space lattice vectors are given by $\bm{g}_{i}$, satisfying $\bm{t}_{i}\cdot\bm{g}_{j}=2\pi \delta_{ij}$.  We indicate symmetry operations using the Seitz notation, $\{\mathcal{R}|\bm{v}\}$, where the translation $\bm{v}$ are expressed in reduced coordinates with respect to the Bravias lattice. The rules for operator multiplication are given by~\cite{Bradley2009Mathematical-Oxford}
\begin{equation}
\begin{aligned}
&\{\mathcal{R}_{1}|\bm{v}_{1}\}\{\mathcal{R}_{2}|\bm{v}_{2}\} = \{\mathcal{R}_{1}\mathcal{R}_{2}|\mathcal{R}_{1}\bm{v}_{2}+\bm{v}_{1}\} \\
&\{\mathcal{R}|\bm{v}\}^{-1} = \{\mathcal{R}^{-1}|-\mathcal{R}^{-1}\bm{v}\} \\
&\{\mathcal{R}|\bm{0}\}\{E|\bm{v}\} = \{E|\mathcal{R}\bm{v}\}\{\mathcal{R}|\bm{0}\} \\
\end{aligned}
\end{equation}
where $E$ is the identity. Since we are interested in spin-$1/2$ particles, a $2\pi$ rotation $\bar{E}$ would give an extra factor of  $-1$. The matrix representations for the symmetry operators can be obtained by using Ref.~\cite{Bradley2009Mathematical-Oxford} (c.f. Table 3.2, Table 3.4, and Table 6.7 in \cite{Bradley2009Mathematical-Oxford}). The coordinates of all high-symmetry points are expressed in fractional coordinates of the reciprocal lattice.

\subsection{Chiral QDP}
\subsubsection{QDP in SGs 92 and 96}\label{ASubSubSec_QDP92&96}
As discussed in the main text, SGs 92 and 96 host a chiral QDP at the high-symmetry point $A$ ($\frac{1}{2}, \frac{1}{2}, \frac{1}{2}$).
Take SG 92 as an example. The little group at $A$ contains three generators: $\tilde{C}_{4z}\equiv\left\{ C_{4z}^{+}\big|00\frac{1}{4}\right\}$, $C_{2;1\bar{1}0}\equiv\left\{ C_{2;1\bar{1}0}\big|\frac{1}{2}\frac{1}{2}\frac{3}{4}\right\}$ and ${\cal T}$.

At $A$, we have the relations
\begin{equation}\label{aeq:q92_alg1}
\tilde{C}_{4z}^{4}=T_{001}\bar{E}=-e^{-i2\pi \tilde{k}_{z}}=1, \qquad
C_{2;1\bar{1}0}^{2}=-1,
\end{equation}
and
\begin{equation}\label{aeq:q92_alg2}
\tilde{C}_{4z}C_{2;1\bar{1}0} = T_{\bar{1}01}\bar{E}C_{2;1\bar{1}0}\tilde{C}_{4z}^{3}
= -e^{i2\pi (-\tilde{k}_{x}+\tilde{k}_{z})}C_{2;1\bar{1}0}\tilde{C}_{4z}^{3}
= -C_{2;1\bar{1}0}\tilde{C}_{4z}^{3},
\end{equation}
where $T_{\bm{v}}$ denotes the lattice translation along $\bm{v}$. $\bar{E}$ is the $2\pi$ spin rotation. The Bloch states at $A$ can be chosen as the eigenstates of $\tilde{C}_{4z}$, which we denote as $|c_{4z}\rangle$ with $c_{4z}\in\{\pm1,\ \pm i\}$ the eigenvalue of $\tilde{C}_{4z}$.
Based on Eq.~(\ref{aeq:q92_alg2}), one finds that
\begin{equation}
\tilde{C}_{4z}C_{2;1\bar{1}0}|\pm1\rangle =-C_{2;1\bar{1}0}\tilde{C}_{4z}^{3}|\pm1\rangle
=\mp C_{2;1\bar{1}0}|\pm1\rangle,
\end{equation}
which indicates that $C_{2;1\bar{1}0}|\pm1\rangle\Rightarrow|\mp1\rangle$ and the two states $|1\rangle$ and $|-1\rangle$ would be degenerate.
In addition, due to ${\cal{T}}^2=-1$, the state $|\pm1\rangle$ and its time-reversal partner ${\cal{T}}|\pm1\rangle$ are linearly independent.
Hence, the four states $\{|1\rangle,\ C_{2;1\bar{1}0}|1\rangle,\ \mathcal{T}|1\rangle,\ \mathcal{T}C_{2;1\bar{1}0}|1\rangle\}$ must be degenerate at the same energy, forming a Dirac point.

The matrix representations of the generators can be expressed in the above quartet basis as
\begin{equation}
\tilde{C}_{4z} = \sigma_{0}\otimes\sigma_{z}, \qquad
C_{2;1\bar{1}0} = i\sigma_{0}\otimes\sigma_{y}, \qquad
\mathcal{T} = i\sigma_{y}\otimes\sigma_{0}\mathcal{K}.
\end{equation}

The Hamiltonian $\mathcal{H}_{\text{eff}}$ is required to be invariant under the symmetry transformations, namely,
\begin{eqnarray}
\tilde{C}_{4z}\mathcal{H}_{\text{eff}}(\bm{k})\tilde{C}_{4z}^{-1} & = & {\mathcal{H}}_{\text{eff}}(-k_{y},k_{x},k_{z}), \label{aeq:q92_S4z} \\
C_{2;1\bar{1}0}{\mathcal{H}}_{\text{eff}}(\bm{k})C_{2;1\bar{1}0}^{-1} & = & {\mathcal{H}}_{\text{eff}}(-k_{y},-k_{x},-k_{z}), \label{aeq:q92_S2b} \\
\mathcal{T}{\mathcal{H}}_{\text{eff}}(\bm{k})\mathcal{T}^{-1} & = & {\mathcal{H}}_{\text{eff}}(-\bm{k}), \label{aeq:q92_T}
\end{eqnarray}
where $\bm{k}$ is measured from the Dirac point.
One notes that from Eq.~(\ref{aeq:q92_S4z}) together with $\tilde{C}_{4z}^{2}=\sigma_{0}\otimes\sigma_{0}$, the Hamiltonian must satisfy
${\mathcal{H}}_{\text{eff}}(\bm{k}) = {\mathcal{H}}_{\text{eff}}(-k_{x},-k_{y},k_{z})$. This clearly eliminates terms which are odd in $k_x$ and $k_y$,  indicating that the Dirac point might be a QDP.

With the constraints Eq.~(\ref{aeq:q92_S4z})-(\ref{aeq:q92_T}), the effective model expanded up to the second order is given as
\begin{eqnarray}
h_{11}^{92}(\bm{k}) & = & c_{1}k_{z}\sigma_{z}+[(c_{2}k_{+}^{2}+c_{3}k_{-}^{2})\sigma_{+} + \text{H.c.}],\\
h_{12}^{92}(\bm{k}) & = & \alpha_{1}k_{z}\sigma_{z}+\alpha_{2}k_{x}k_{y}\sigma_{y},
\end{eqnarray}
and
\begin{eqnarray}
h_{22}^{92}(\bm{k})=h_{11}^{92*}(-\bm{k})
\end{eqnarray}
is the time-reversed partner of $h_{11}^{92}$. Here, $c_{i}$ are real parameters and $\alpha_{i}$ are complex parameters. This effective model confirms that the Dirac point is a QDP, with linear band splitting along $k_z$ and quadratic splitting in the $k_x$-$k_y$ plane. The diagonal blocks $h_{11}^{92}$ and its time-reversed partner $h_{22}^{92}$ each corresponds to a double Weyl point, and they share the same topological charge (Chern number) of $2\rm{sgn}(|c_{2}|-|c_{3}|)$. Therefore, the Dirac point is chiral and has topological charge $\mathcal{C}=\pm 4$.

Similar analysis applies for SG~96, whose $k\cdot p$ model $\mathcal{H}_{\text{eff}}^{96}$ takes the same form with SG~92, that is
\begin{equation}
\mathcal{H}_{\text{eff}}^{96}(\bm{k}) = \mathcal{H}_{\text{eff}}^{92}(\bm{k}).
\end{equation}

\subsection{QDP at non-TRIM points}
\subsubsection{QDP in SG 142}\label{ASubSubSec_QDP142}
SG~142 hosts a QDP at the high-symmetry point $P$ ($\frac{1}{4}, \frac{1}{4}, \frac{1}{4}$) which is not a TRIM point.
The little group at $P$ contains three generators: $S_{4z}\equiv\left\{ S_{4z}^{+}\big|\frac{1}{2}\frac{1}{2}\frac{1}{2}\right\}$, $\tilde{M}_{1\bar{1}0}\equiv\left\{ M_{1\bar{1}0}\big|10\frac{1}{2}\right\}$, and ${\cal PT}$.
The following relations are satisfied at $P$:
\begin{equation}\label{aeq:q142_alg1}
S_{4z}^{4}=-1, \qquad
\tilde{M}_{1\bar{1}0}^{2}=T_{111}\bar{E}=-e^{-i2\pi (\tilde{k}_{x}+\tilde{k}_{y}+\tilde{k}_{z})}=-i,
\end{equation}
and
\begin{eqnarray}
	S_{4z}\tilde{M}_{1\bar{1}0} & = & T_{\bar{1}\bar{1}\bar{1}}\bar{E}\tilde{M}_{1\bar{1}0}S_{4z}^{3}
	= -e^{i2\pi (\tilde{k}_{x}+\tilde{k}_{y}+\tilde{k}_{z})}\tilde{M}_{1\bar{1}0}S_{4z}^{3}
	= i\tilde{M}_{1\bar{1}0}S_{4z}^{3}, \label{aeq:q142_alg2} \\
	S_{4z}(\mathcal{PT}) & = &  T_{111}(\mathcal{PT})S_{4z}
	= e^{-i2\pi (\tilde{k}_{x}+\tilde{k}_{y}+\tilde{k}_{z})}(\mathcal{PT})S_{4z}
	= i(\mathcal{PT})S_{4z}, \label{aeq:q142_alg3} \\
	\tilde{M}_{1\bar{1}0}(\mathcal{PT}) & = & T_{201}(\mathcal{PT})\tilde{M}_{1\bar{1}0}
	= e^{-i2\pi (2\tilde{k}_{x}+\tilde{k}_{z})}(\mathcal{PT})\tilde{M}_{1\bar{1}0}
	= i(\mathcal{PT})\tilde{M}_{1\bar{1}0}. \label{aeq:q142_alg4}
\end{eqnarray}
The Bloch states at $P$ can be chosen as the eigenstates of $S_{4z}$, which we denote as $|s_{4z}\rangle$ with $s_{4z}\in\{\pm e^{i\pi/4},\ \pm e^{- i\pi/4}\}$ the eigenvalue of $S_{4z}$.
Based on Eq.~(\ref{aeq:q142_alg2}), one finds that
\begin{equation}
S_{4z}\tilde{M}_{1\bar{1}0}|\pm e^{i\pi/4}\rangle=i\tilde{M}_{1\bar{1}0}S_{4z}^{3}|\pm e^{i\pi/4}\rangle
= \mp e^{i\pi/4}\tilde{M}_{1\bar{1}0}|\pm e^{i\pi/4}\rangle,
\end{equation}
which indicates that $\tilde{M}_{1\bar{1}0}|\pm e^{i\pi/4}\rangle\Rightarrow|\mp e^{i\pi/4}\rangle$ and the two states $|e^{i\pi/4}\rangle$ and $\tilde{M}_{1\bar{1}0}|e^{i\pi/4}\rangle$ would be degenerate. In addition, since $({\cal{PT}})^2=-1$, the state $|\pm e^{i\pi/4}\rangle$ and its Kramers partner ${\cal{PT}}|\pm e^{i\pi/4}\rangle$ are linearly independent.
One finds that $|s_{4z}\rangle$ and $\tilde{M}_{1\bar{1}0}|s_{4z}\rangle$ share opposite $s_{4z}$ while $|s_{4z}\rangle$ and ${\cal{PT}} |s_{4z}\rangle$ have same $s_{4z}$, as 
\begin{equation}
S_{4z}(\mathcal{PT})|\pm e^{i\pi/4}\rangle = i(\mathcal{PT})S_{4z}|\pm e^{i\pi/4}\rangle
= \pm e^{i\pi/4}(\mathcal{PT})|\pm e^{i\pi/4}\rangle.
\end{equation}
Hence,
the four states $\{|e^{i\pi/4}\rangle,\ \tilde{M}_{1\bar{1}0}|e^{i\pi/4}\rangle,\ \mathcal{PT}|e^{i\pi/4}\rangle,\ (\mathcal{PT})\tilde{M}_{1\bar{1}0}|e^{i\pi/4}\rangle\}$ must be linearly independent and degenerate at the same energy, forming a Dirac point.

The matrix representations of the generators can be expressed in the above quartet basis as
\begin{equation}
S_{4z} = e^{i\pi/4}\sigma_{0}\otimes\sigma_{z}, \qquad
\tilde{M}_{1\bar{1}0} = e^{i3\pi/4}(\frac{1}{\sqrt{2}}\sigma_{0}\otimes\sigma_{y} - \frac{1}{\sqrt{2}}\sigma_{z}\otimes\sigma_{x}), \qquad
\mathcal{PT} = i\sigma_{y}\otimes\sigma_{0}\mathcal{K}
\end{equation}
and the symmetry constrains on Hamiltonian are
\begin{eqnarray}
S_{4z}\mathcal{H}_{\text{eff}}(\bm{k})S_{4z}^{-1} & = & \mathcal{H}_{\text{eff}}(-k_{y},k_{x},-k_{z}), \label{aeq:q142_S4z}\\
\tilde{M}_{1\bar{1}0}\mathcal{H}_{\text{eff}}(\bm{k})\tilde{M}_{1\bar{1}0}^{-1} & = & \mathcal{H}_{\text{eff}}(k_{y},k_{x},k_{z}),\label{aeq:q142_S2b}\\
(\mathcal{PT})\mathcal{H}_{\text{eff}}(\bm{k})(\mathcal{PT})^{-1} & = & \mathcal{H}_{\text{eff}}(\bm{k}).\label{aeq:q142_PT}
\end{eqnarray}
One notes that from Eq. (\ref{aeq:q142_S4z}) together with $S_{4z}^{2}=i\sigma_{0}\otimes\sigma_{0}$, the Hamiltonian must satisfy $\mathcal{H}_{\text{eff}}(-k_{x},-k_{y},k_{z})=\mathcal{H}_{\text{eff}}(\bm{k})$, which clearly eliminates terms which are odd in $k_{x}$ and $k_{y}$. This indicates that the Dirac point might be a QDP.

Under the symmetry constraint in Eq.~(\ref{aeq:q142_S4z})-(\ref{aeq:q142_PT}), the effective model expanded up to the second order is given as
\begin{eqnarray}
h_{11}^{142}(\bm{k}) & = & e^{-i\pi/4}(ic_{1} k_{z} + c_{2}k_{+}^{2}+c_{3}k_{-}^{2})\sigma_{+} + \rm{H.c.}, \label{aeq:q142_H11} \\
h_{12}^{142}(\bm{k}) & = & (\alpha_{1}k_{z}+\alpha_{2}k_{x}k_{y})\sigma_{y}, \label{aeq:q142_H12}
\end{eqnarray}
and
\begin{equation}
h_{22}^{142}(\bm{k})=h_{11}^{142*}(\bm{k}). \label{aeq:q142_H22}
\end{equation}
This effective model confirms that the Dirac point is a QDP, with linear band splitting along $k_z$ and quadratic splitting in the $k_x$-$k_y$ plane.

\subsubsection{QDP of SG 228}
SG~228 hosts a QDP at the high-symmetry point $W$ ($\frac{1}{2}, \frac{1}{4}, \frac{3}{4}$) which is not a TRIM point.
The little group at $W$ contains three generators contains three generators: $S_{4x}\equiv\left\{ S_{4x}^{+}\big|\frac{1}{2}\frac{1}{2}\frac{1}{2}\right\}$, $\tilde{M}_{z}\equiv\left\{ M_{z}\big|\frac{3}{4}\frac{3}{4}\frac{3}{4}\right\}$ and $\mathcal{PT}\equiv\mathcal{T}\left\{ \mathcal{P}\big|\frac{3}{4}\frac{3}{4}\frac{3}{4}\right\}$.
The following relations are satisfied at $W$:
\begin{equation}\label{aeq:q228_alg1}
S_{4x}^{4}=-1, \qquad
\tilde{M}_{z}^{2}=T_{003}\bar{E}= -e^{-i2\pi (3\tilde{k}_{z})}=i
\end{equation}
and
\begin{eqnarray}
S_{4x}\tilde{M}_{z} & = & T_{20\bar{3}}\bar{E}\tilde{M}_{z}S_{4x}^{3}
= -e^{-i2\pi (2\tilde{k}_{x}-3\tilde{k}_{z})}\tilde{M}_{z}S_{4x}^{3}
= -i\tilde{M}_{z}S_{4x}^{3}, \label{aeq:q228_alg2} \\
S_{4x}(\mathcal{PT}) & = & T_{11\bar{2}}(\mathcal{PT})S_{4z}
= e^{-i2\pi (\tilde{k}_{x}+\tilde{k}_{y}-2\tilde{k}_{z})}(\mathcal{PT})S_{4z}
= -i(\mathcal{PT})S_{4z}, \label{aeq:q228_alg3} \\
\tilde{M}_{z}(\mathcal{PT}) & = & T_{003}(\mathcal{PT})\tilde{M}_{z}
= e^{-i2\pi (3\tilde{k}_{z})}(\mathcal{PT})\tilde{M}_{z}
= -i(\mathcal{PT})\tilde{M}_{z}. \label{aeq:q228_alg4}
\end{eqnarray}

The Bloch states at $P$ can be chosen as the eigenstates of $S_{4x}$, which we denote as $|s_{4x}\rangle$ with $s_{4x}\in\{\pm e^{i\pi/4},\ \pm e^{- i\pi/4}\}$ the eigenvalue of $S_{4x}$.
Based on Eq.~(\ref{aeq:q228_alg2}), one finds that
\begin{equation}
S_{4x}\tilde{M}_{z}|\pm e^{-i\pi/4}\rangle=-i\tilde{M}_{z}S_{4x}^{3}|\pm e^{-i\pi/4}\rangle
= \mp e^{-i\pi/4}\tilde{M}_{z}|\pm e^{-i\pi/4}\rangle,
\end{equation}
which indicates that $\tilde{M}_{z}|\pm e^{-i\pi/4}\rangle\Rightarrow|\mp e^{-i\pi/4}\rangle$ and the two states $|e^{-i\pi/4}\rangle$ and $|-e^{-i\pi/4}\rangle$ would be degenerate. Also, since $({\cal{PT}})^2=-1$, the state $|\pm e^{-i\pi/4}\rangle$ and its Kramers partner ${\cal{PT}}|\pm e^{-i\pi/4}\rangle$ are linearly independent. Following the commutation relation in Eq.~(\ref{aeq:q228_alg3}), one finds that
\begin{equation}
S_{4x}(\mathcal{PT})|\pm e^{-i\pi/4}\rangle = -i(\mathcal{PT})S_{4x}|\pm e^{-i\pi/4}\rangle
= \pm e^{-i\pi/4}(\mathcal{PT})|\pm e^{-i\pi/4}\rangle,
\end{equation}
which indicates that $|\pm e^{-i\pi/4}\rangle$ and ${\cal{PT}}|\pm e^{-i\pi/4}\rangle$ share the same $s_{4x}$. Since ${\cal{PT}}|\pm e^{-i\pi/4}\rangle$ and $\tilde{M}_{z}|\pm e^{-i\pi/4}\rangle$ have the opposite $s_{4z}$, the four states $\{|e^{-i\pi/4}\rangle,\ \tilde{M}_{z}|e^{-i\pi/4}\rangle,\ \mathcal{PT}|e^{-i\pi/4}\rangle,\ (\mathcal{PT})\tilde{M}_{z}|e^{-i\pi/4}\rangle\}$ must be linearly independent and degenerate at the same energy, forming a Dirac point.

The matrix representations of the generators can be expressed in the above quartet basis as
\begin{equation}
S_{4x} = e^{-i\pi/4}\sigma_{0}\otimes\sigma_{z}, \qquad
\tilde{M}_{z} = e^{i\pi/4}(\frac{1}{\sqrt{2}}\sigma_{0}\otimes\sigma_{y} + \frac{1}{\sqrt{2}}\sigma_{z}\otimes\sigma_{x}), \qquad
\mathcal{PT} = i\sigma_{y}\otimes\sigma_{0}\mathcal{K},
\end{equation}
and the symmetry constrains on Hamiltonian are
\begin{eqnarray}
S_{4x}\mathcal{H}_{\text{eff}}(\bm{k})S_{4x}^{-1} & = & \mathcal{H}_{\text{eff}}(-k_{x},-k_{z},k_{y}), \label{aeq:q228_S4x}\\
\tilde{M}_{z}\mathcal{H}_{\text{eff}}(\bm{k})\tilde{M}_{z}^{-1} & = & \mathcal{H}_{\text{eff}}(k_{x},k_{y},-k_{z}),\label{aeq:q228_My}\\
\mathcal{PT}\mathcal{H}_{\text{eff}}(\bm{k})(\mathcal{PT})^{-1} & = & \mathcal{H}_{\text{eff}}(\bm{k}).\label{aeq:q228_PT}
\end{eqnarray}
where $\bm{k}$ is measured from the Dirac point. One notes that from Eq. (\ref{aeq:q228_S4x}) together with $S_{4x}^{2}=-i\sigma_{0}\otimes\sigma_{0}$, the Hamiltonian must satisfy $\mathcal{H}_{\text{eff}}(k_{x},-k_{y},-k_{z})=\mathcal{H}_{\text{eff}}(\bm{k})$, which eliminates terms which are odd in $k_{y}$ and $k_{z}$. This indicates that the Dirac point might be a QDP.

Under the symmetry constraint in Eq.~(\ref{aeq:q228_S4x})-(\ref{aeq:q228_PT}), the effective model expanded up to the second order is given as
\begin{eqnarray}
h_{11}^{228}(\bm{k}) & = & e^{i\pi/4}(ic_{1} k_{x} + c_{2}k_{+}^{2}+c_{3}k_{-}^{2})\sigma_{+} + \rm{H.c.},\label{aeq:q228_H11}\\
h_{12}^{228}(\bm{k}) & = & [\alpha_{1}k_{x}+\alpha_{2}k_{0\bar{1}1}k_{0\bar{1}\bar{1}}]\sigma_{y},\label{aeq:q228_H12}
\end{eqnarray}
and
\begin{equation}
h_{22}^{228}(\bm{k})=h_{11}^{228*}(\bm{k}).\label{aeq:q228_H22}
\end{equation}
where $k_{0\bar{1}1}=-k_{y}+k_{z}$, $k_{0\bar{1}\bar{1}}=-k_{y}-k_{z}$, and $k_{\pm} = k_{0\bar{1}1}\pm ik_{0\bar{1}\bar{1}}$.
This effective model confirms that the Dirac point is a QDP, with linear band splitting along $k_x$ and quadratic splitting in the $k_y$-$k_z$ plane.
Notably, the effective Hamiltonian has the same form to the model of SG~142 [see Eq.~(\ref{aeq:q142_H11})-(\ref{aeq:q142_H22})], after an unitary transformation and a coordinate transformation,
\begin{equation}
\mathcal{H}_{\text{eff}}^{228}(k_{0\bar{1}1},k_{0\bar{1}\bar{1}},k_{x}) \sim U \mathcal{H}_{\text{eff}}^{142}(k_{x},k_{y},k_{z})U^{-1}
\end{equation}
where
\begin{equation}
U=\left[\begin{array}{cccc}
1 & 0 & 0 & 0\\
0 & -i & 0 & 0\\
0 & 0 & 1 & 0\\
0 & 0 & 0 & i
\end{array}\right]
\end{equation}

\subsection{CDP without inversion symmetry}
\subsubsection{CDP in SG 184}
SG 184 hosts a CDP at the high-symmetry point $A$ ($0,0,\frac{1}{2}$). The little group at $A$ contains three generators: $C_{6z}\equiv\left\{ C_{6z}^{+}\big|000\right\}$, $\tilde{M}_{y}\equiv\left\{ M_{y}\big|00\frac{1}{2}\right\}$, and ${\cal T}$. The following relations are satisfied at $A$:
\begin{equation}\label{aeq:c184_alg1}
\begin{aligned}
C_{6z}^{6} = -1, \qquad
\tilde{M}_{y}^{2} = T_{001}\bar{E} = -e^{i2\pi \tilde{k}_{z}} = 1,
\end{aligned}
\end{equation}
and
\begin{equation}\label{aeq:c184_alg2}
C_{6z}\tilde{M}_{y} = -\tilde{M}_{y}C_{6z}^{5}.
\end{equation}
The Bloch states at $A$ can be chosen as the eigenstates of $C_{6z}$, which we denotes as $|c_{6z}\rangle$ with the $C_{6z}$ eigenvalue $c_{6z}\in \{\pm e^{i\pi/6}, \pm i, \pm e^{-i\pi/6}\}$. Based on Eq.~(\ref{aeq:c184_alg2}), one finds that
\begin{equation}
C_{6z}\tilde{M}_{y} |\pm i\rangle = -\tilde{M}_{y}C_{6z}^{5}|\pm i\rangle
=\mp i\tilde{M}_{y}|\pm i\rangle,
\end{equation}
which indicates that $\tilde{M}_{y} |\pm i\rangle \Rightarrow |\mp i\rangle$ and the two states $|i\rangle$ and $|-i\rangle$ would be degenerate. In addition, due to $\mathcal{T}^{2}=-1$, the states $|\pm i\rangle$ and its time-reversal partner $\mathcal{T}|\pm i\rangle$ are linearly independent. Notice  that $\tilde{M}_{y} |c_{6z}\rangle$ and $\mathcal{T}|c_{6z}\rangle$ also are orthogonal to each other. To prove this, one can assume that $\tilde{M}_{y} |c_{6z}\rangle$ and $\mathcal{T}|c_{6z}\rangle$ are the same state, then one generally has $\tilde{M}_{y} |c_{6z}\rangle = e^{i\phi}\mathcal{T}|c_{6z}\rangle$.
Since $\tilde{M}_{y}^{2}=1$, one would have
\begin{equation}
|c_{6z}\rangle = \tilde{M}_{y}^{2}|c_{6z}\rangle
= e^{i\phi}\mathcal{T}\tilde{M}_{y}|c_{6z}\rangle
= e^{i\phi}\mathcal{T}e^{i\phi}\mathcal{T}|c_{6z}\rangle
= \mathcal{T}^{2}|c_{6z}\rangle = -|c_{6z}\rangle,
\end{equation}
which apparently is contradictory, meaning that $\tilde{M}_{y} |c_{6z}\rangle$ and $\mathcal{T}|c_{6z}\rangle$ are two linearly independent states.
Therefore, the four states $\{|i\rangle, \mathcal{T}|i\rangle, \tilde{M}_{y}|i\rangle, \mathcal{T}\tilde{M}_{y}|i\rangle\}$ must be distinct, and they degenerate at the same energy, forming a Dirac point.

The matrix representations of the generators in the above quartet basis can be expressed as
\begin{equation}
\begin{aligned}
C_{6z} = i\sigma_{z}\otimes\sigma_{z}, \qquad
\tilde{M}_{y} = \sigma_{x}\otimes\sigma_{0}, \qquad
\mathcal{T} = \sigma_{0}\otimes i\sigma_{y} \mathcal{K},
\end{aligned}
\end{equation}
and the symmetry constraints on Hamiltonian are
\begin{eqnarray}
C_{6z}{\cal H}(\bm{k})C_{6z}^{-1} & = & {\cal H}(R_{6z}\bm{k}), \label{aeq:c184_C6z} \\
\tilde{M}_{y}{\cal H}(\bm{k})\tilde{M}_{y}^{-1} & = & {\cal H}(k_{x},-k_{y},k_{z}), \label{aeq:c184_My}\\
\mathcal{T}{\cal H}(\bm{k})\mathcal{T}^{-1} & = & {\cal H}(-\bm{k}). \label{aeq:c184_T}
\end{eqnarray}
One notes that from Eq. (\ref{aeq:c184_C6z}) together with $C_{6z}^2=\sigma_{0}\otimes\sigma_{0}$, the Hamiltonian must satisfy $\mathcal{H}_{\text{eff}}(\bm{k})=\mathcal{H}_{\text{eff}}(R_{3z}\bm{k})$ which eliminates both linear and quadratic terms in $k_{x}$ and $k_{y}$, but the cubic term can be present. This indicates that the Dirac point is a CDP.

Under the symmetry constraint in Eq.~(\ref{aeq:c184_C6z})-(\ref{aeq:c184_T}), the effective model expanded up to the third order is given as
\begin{eqnarray}
h_{11}^{184}(\bm{k}) & = & f(\bm{k})k_{z}\sigma_{z}+[(\alpha_{1}k_{+}^{3}+\alpha_{2}k_{-}^{3})\sigma_{+} + \text{H.c.}], \\
h_{12}^{184}(\bm{k}) & = & (c_{4}k_{+}^{3} + c_{5}k_{-}^{3})\sigma_{0} + [g(\bm{k})k_{z}\sigma_{+} + \text{H.c.}],
\end{eqnarray}
and
\begin{equation}
h_{22}^{184}(\bm{k})=h_{11}^{184}(k_{x},-k_{y},k_{z}).
\end{equation}
where $f(\bm{k})=c_{1}+c_{2}(k_{x}^{2}+k_{y}^{2})+c_{3}k_{z}^{2}$ and $g(\bm{k})=\alpha_{3}+\alpha_{4}(k_{x}^{2}+k_{y}^{2})+\alpha_{5}k_{z}^{2}$.
This effective model confirms that the Dirac point is a CDP, with linear band splitting along $k_z$ and cubic splitting in the $k_x$-$k_y$ plane.

\subsubsection{CDP in SGs 185 and 186}
Both SG 185 and SG 186 host a CDP at the high-symmetry point $A$ ($0,0,\frac{1}{2}$). Take SG 185 as an example. The little group at $A$ contains three generators:
$\tilde{C}_{6z}\equiv\left\{ C_{6z}^{+}\big|00\frac{1}{2}\right\} $, $M_{y}\equiv\left\{ M_{y}\big|000\right\} $, and $\mathcal{T}$. The following relations are satisfied at $A$:
\begin{equation}\label{aeq:c185_alg1}
\tilde{C}_{6z}^{6} = T_{003}\bar{E} = -e^{i2\pi (3\tilde{k}_{z})} = 1, \qquad
M_{y}^{2} = -1
\end{equation}
and
\begin{equation}\label{aeq:c185_alg2}
\tilde{C}_{6z}M_{y} = T_{00\bar{2}}\bar{E}M_{y}\tilde{C}_{6z}^{5}
= -e^{i2\pi (2\tilde{k}_{z})}M_{y}\tilde{C}_{6z}^{5}
= -M_{y}\tilde{C}_{6z}^{5}.
\end{equation}
The Bloch states at $A$ can be chosen as the eigenstates of $C_{6z}$, which we denotes as $|c_{6z}\rangle$ with the $C_{6z}$ eigenvalue $c_{6z}\in \{\pm e^{i\pi/3}, \pm 1, \pm e^{-i\pi/3}\}$. Based on Eq.~(\ref{aeq:c185_alg2}), one finds that
\begin{equation}
\tilde{C}_{6z}M_{y} |\pm 1\rangle = -M_{y}\tilde{C}_{6z}^{5}|\pm 1\rangle
= (\mp 1)\tilde{M}_{y}|\pm 1\rangle,
\end{equation}
which indicates that $M_{y} |\pm 1\rangle \Rightarrow |\mp 1\rangle$ and the two states $|1\rangle$ and $|-1\rangle$ would be degenerate. In addition, due to $\mathcal{T}^{2}=-1$, the states $|\pm 1\rangle$ and its time-reversal partner $\mathcal{T}|\pm 1\rangle$ are linearly independent. Hence, the four states $\{|1\rangle, M_{y}|1\rangle, \mathcal{T}|1\rangle, \mathcal{T}M_{y}|1\rangle\}$ must be degenerate at the same energy, forming a Dirac point.

The matrix representations of the generators in the above quartet basis can be expressed as
\begin{equation}
\begin{aligned}
\tilde{C}_{6z} = \sigma_{0}\otimes\sigma_{z}, \qquad
M_{y} = i\sigma_{0}\otimes\sigma_{y}, \qquad
\mathcal{T} = i\sigma_{y} \otimes \sigma_{0} \mathcal{K},
\end{aligned}
\end{equation}
and the symmetry constraints on Hamiltonian are
\begin{eqnarray}
\tilde{C}_{6z}{\cal H}(\bm{k})\tilde{C}_{6z}^{-1} & = & {\cal H}(R_{6z}\bm{k}), \label{aeq:c185_C6z} \\
M_{y}{\cal H}(\bm{k})M_{y}^{-1} & = & {\cal H}(k_{x},-k_{y},k_{z}), \label{aeq:c185_My}\\
\mathcal{T}{\cal H}(\bm{k})\mathcal{T}^{-1} & = & {\cal H}(-\bm{k}). \label{aeq:c185_T}
\end{eqnarray}
One notes that from Eq. (\ref{aeq:c185_C6z}) together with $C_{6z}^2=\sigma_{0}\otimes\sigma_{0}$, the Hamiltonian must satisfy $\mathcal{H}_{\text{eff}}(\bm{k})=\mathcal{H}_{\text{eff}}(R_{3z}\bm{k})$ which eliminates both linear and quadratic terms in $k_{x}$ and $k_{y}$, but the cubic term can be present. This indicates that the Dirac point would be a CDP.

Under the symmetry constraint in Eq.~(\ref{aeq:c185_C6z})-(\ref{aeq:c185_T}), the effective model expanded up to the third order is given as
\begin{eqnarray}
h_{11}^{185}(\bm{k}) & = & f(\bm{k})k_{z}\sigma_{0}+[i(c_{4}k_{+}^{3}+c_{5}k_{-}^{3})\sigma_{+} + \text{H.c.}],\\
h_{12}^{185}(\bm{k}) & = & g(\bm{k})k_{z}\sigma_{0} + \alpha_{4}(k_{+}^{3}-k_{-}^{3})\sigma_{x},
\end{eqnarray}
and
\begin{equation}
h_{22}^{185}(\bm{k})=h_{11}^{185*}(-\bm{k}).
\end{equation}
where $f(\bm{k})=c_{1}+c_{2}(k_{x}^{2}+k_{y}^{2})+c_{3}k_{z}^{2}$ and $g(\bm{k})=\alpha_{1}+\alpha_{2}(k_{x}^{2}+k_{y}^{2})+\alpha_{3}k_{z}^{2}$. 
This effective model confirms that the Dirac point is a CDP, with linear band splitting along $k_z$ and cubic splitting in the $k_x$-$k_y$ plane.

For SG 186, instead of $M_{y}$, we have $M_{x}=i\sigma_{0}\otimes\sigma_{y}$, and the effective model of the CDP is related with the model of SG 185 by a coordinate transformation, which reads
\begin{equation}
\mathcal{H}_{\text{eff}}^{186}(k_{x},k_{y},k_{z}) = \mathcal{H}_{\text{eff}}^{185}(-k_{y},k_{x},k_{z})
\end{equation}

\subsection{CDP with inversion symmetry}
\subsubsection{CDP in SGs 163, 165, 167, 226, and 228}
Take SG 163 as an example. SG 163 hosts a CDP at high-symmetry point $A$ ($0,0,\frac{1}{2}$). The little group at $A$ contains four generators: $C_{3z}\equiv\left\{C_{3z}^{+}\big|000\right\} $, $C_{2y}\equiv\left\{ C_{2y}\big|00\frac{1}{2}\right\}$,
$\mathcal{P}$ and $\mathcal{T}$. The following relations are satisfied at $A$:
\begin{equation}\label{aeq:c163_alg1}
C_{3z}^{3} =-1, \qquad C_{2y}^{2} = -1, \qquad \mathcal{P}^{2} = 1,
\end{equation}
and
\begin{eqnarray}
C_{3z}C_{2y} & = & -C_{2y}C_{3z}^{2}, \label{aeq:c163_alg2} \\
\mathcal{P}C_{3z} & = & C_{3z}\mathcal{P}, \label{aeq:c163_alg3} \\
\mathcal{P}C_{2y} = T_{00\bar{1}}C_{2y}\mathcal{P}
& = & e^{i2\pi \tilde{k}_{z}}C_{2y}\mathcal{P}
= -C_{2y}\mathcal{P}. \label{aeq:c163_alg4}
\end{eqnarray}

The Bloch states at $A$ can be chosen as the eigenstates of $\mathcal{P}$, which we denote as $|p\rangle$ with the $\mathcal{P}$ eigenvalue $p\in \{\pm 1\}$. Based on Eq.~(\ref{aeq:c163_alg4}), one finds that
\begin{equation}
\mathcal{P}C_{2y} |\pm 1\rangle = -C_{2y}\mathcal{P}|\pm 1\rangle
=(\mp 1)C_{2y}|\pm 1\rangle,
\end{equation}
which indicates that $C_{2y} |\pm 1\rangle \Rightarrow |\mp 1\rangle$ and the two states $|1\rangle$ and $|-1\rangle$ would be degenerate. In addition, due to $\mathcal{T}^{2}=-1$, the states $|\pm 1\rangle$ and its time-reversal partner $\mathcal{T}|\pm 1\rangle$ are linearly independent. Hence, the four states $\{|1\rangle, C_{2y}|1\rangle, \mathcal{T}|1\rangle, \mathcal{T}C_{2y}|1\rangle\}$ must be degenerate at the same energy, forming a Dirac point.

Moreover, since $[\mathcal{P},C_{3z}]=0$, the Bloch states at $A$ also can be chosen as the eigenstates of $C_{3z}$ with the $C_{3z}$ eigenvalue $c_{3z}\in \{e^{\pm i\pi/3}, -1\}$ and the states thus can be denoted as $|c_{3z}, p\rangle$. We choose the degenerate quartet states $\{|-1, 1\rangle, C_{2y}|-1, 1\rangle, \mathcal{T}|-1, 1\rangle, \mathcal{T}C_{2y}|-1, 1\rangle\}$ as our basis, so that the matrix representations of the generators can be expressed as
\begin{equation}
C_{3z} = -\sigma_{0}\otimes\sigma_{0}, \qquad
C_{2y} = i\sigma_{0}\otimes\sigma_{y}, \qquad
\mathcal{P} = \sigma_{0}\otimes\sigma_{z}, \qquad
\mathcal{T} = i\sigma_{y}\otimes\sigma_{0}  \mathcal{K},
\end{equation}
and the symmetry constraints on Hamiltonian are
\begin{eqnarray}
C_{3z}{\cal H}(\bm{k})C_{3z}^{-1} & = & {\cal H}(R_{3z}\bm{k}), \label{aeq:c163_C3z} \\
C_{2y}{\cal H}(\bm{k})C_{2y}^{-1} & = & {\cal H}(-k_{x},k_{y},-k_{z}), \label{aeq:c163_C2y}\\
\mathcal{P}{\cal H}(\bm{k})\mathcal{P}^{-1} & = & {\cal H}(-\bm{k}). \label{aeq:c163_P} \\
\mathcal{T}{\cal H}(\bm{k})\mathcal{T}^{-1} & = & {\cal H}(-\bm{k}). \label{aeq:c163_T}
\end{eqnarray}
One notes that from Eq. (\ref{aeq:c163_C3z}) together with $C_{3z}=-\sigma_{0}\otimes\sigma_{0}$, the Hamiltonian must satisfy $\mathcal{H}_{\text{eff}}(\bm{k})=\mathcal{H}_{\text{eff}}(R_{3z}\bm{k})$ which eliminates both linear and quadratic terms in $k_{x}$ and $k_{y}$, but the cubic term can be present. This indicates that the Dirac point would be a CDP.

Under the symmetry constraint in Eq.~(\ref{aeq:c163_C3z})-(\ref{aeq:c163_T}), the effective model expanded up to the third order is given as
\begin{eqnarray}
h_{11}^{163}(\bm{k}) & = & [f(\bm{k})k_{z} + c_{4}k_{+}^{3} + c_{5}k_{-}^{3}]\sigma_{+} + \text{H.c.},\\
h_{12}^{163}(\bm{k}) & = & [g(\bm{k})k_{z}+\alpha_{4}(k_{+}^{3}+k_{-}^{3})]\sigma_{x},
\end{eqnarray}
and
\begin{equation}
h_{22}^{163}(\bm{k})=h_{11}^{163*}(-\bm{k}).
\end{equation}
where $f(\bm{k})=c_{1}+c_{2}(k_{x}^{2}+k_{y}^{2})+c_{3}k_{z}^{2}$ and $g(\bm{k})=\alpha_{1}+\alpha_{2}(k_{x}^{2}+k_{y}^{2})+\alpha_{3}k_{z}^{2}$.
This effective model confirms that the Dirac point is a CDP, with linear band splitting along $k_z$ and cubic splitting in the $k_x$-$k_y$ plane.

\ \

For SG 165, instead of $C_{2y}$, we have $C_{2x}=i\sigma_{0}\otimes\sigma_{y}$, and the effective model of the CDP is related with the model of SG 163 by a coordinate transformation, which reads
\begin{equation}\label{aeq:c165_Hamiltonian}
\mathcal{H}_{\text{eff}}^{165}(k_{x},k_{y},k_{z}) = \mathcal{H}_{\text{eff}}^{163}(-k_{y},k_{x},k_{z})
\end{equation}

For SG 167, the CDP is located at high-symmetry point $Z$ ($\frac{1}{2},\frac{1}{2},\frac{1}{2}$) and the generators of the little group at $Z$ include $\left\{C_{3z}^{+}\big|000\right\} $, $\left\{ C_{2y}\big|\frac{1}{2}\frac{1}{2}\frac{1}{2}\right\}$, $\mathcal{P}$ and $\mathcal{T}$. We notice that the algebra at $Z$ shares the same feature with that at $A$ in SG 163. Hence, the effective model of the CDP has the same form with the model of SG 163, that is,
\begin{equation}\label{aeq:c167_Hamiltonian}
\mathcal{H}_{\text{eff}}^{167}(k_{x},k_{y},k_{z}) = \mathcal{H}_{\text{eff}}^{163}(k_{x},k_{y},k_{z})
\end{equation}

For SGs 226 and 228, the CDP is located at high-symmetry point $L$ ($\frac{1}{2},\frac{1}{2},\frac{1}{2}$). Different from SG 163, the orientation of $C_{3}$ rotation axis is along the $(111)$ direction, and a twofold rotation axis which is perpendicular to the $C_{3}$ axis is correspondingly along the $(1\bar{1}0)$ direction. We notice that the algebra at $L$ in SGs 226 and 228 shares the same feature with that at $A$ in SG 163. Hence, the effective model of the CDP is related with the model of SG 163 only by a coordinate transformation, which reads
\begin{equation}\label{aeq:c226&228_Hamiltonian}
\mathcal{H}_{\text{eff}}^{226,228}(k_{x},k_{y},k_{z}) = \mathcal{H}_{\text{eff}}^{163}(k_{\bar{1}\bar{1}2},k_{1\bar{1}0},k_{111})
\end{equation}

\subsubsection{CDP in SG 192}
SG 192 hosts a CDP at the high-symmetry point $A$ ($0,0,\frac{1}{2}$). The little group at $A$ contains four generators: $C_{6z}\equiv\left\{ C_{6z}^{+}\big|000\right\} $, $\tilde{M}_{y}\equiv\left\{ M_{y}\big|00\frac{1}{2}\right\} $,
$\mathcal{P}$, and $\mathcal{T}$. The following relations are satisfied at $A$:
\begin{equation}\label{aeq:c192_alg1}
C_{6z}^{6} = -1, \qquad \mathcal{P}^{2} = 1, \qquad
\tilde{M}_{y}^{2} = T_{001}\bar{E} = -e^{-i2\pi \tilde{k}_{z}} = 1
\end{equation}
and
\begin{eqnarray}
C_{6z}\tilde{M}_{y} & = & -\tilde{M}_{y}C_{6z}^{5}, \label{aeq:c192_alg2} \\
C_{6z}\mathcal{P} & = & C_{6z}\mathcal{P}, \label{aeq:c192_alg3} \\
\mathcal{P}\tilde{M}_{y} = T_{00\bar{1}}\tilde{M}_{y}\mathcal{P}
& = & e^{i2\pi \tilde{k}_{z}}\tilde{M}_{y}\mathcal{P}
=-\tilde{M}_{y}\mathcal{P}. \label{aeq:c192_alg4}
\end{eqnarray}

The Bloch states at $A$ can be chosen as the eigenstates of $\mathcal{P}$, which we denote as $|p\rangle$ with the $\mathcal{P}$ eigenvalue $p\in \{\pm 1\}$. Based on Eq.~(\ref{aeq:c192_alg4}), one finds that
\begin{equation}
\mathcal{P}\tilde{M}_{y}|\pm 1\rangle = -\tilde{M}_{y}\mathcal{P}|\pm 1\rangle
=(\mp 1)\tilde{M}_{y}|\pm 1\rangle,
\end{equation}
which indicates that $\tilde{M}_{y}|\pm 1\rangle \Rightarrow |\mp 1\rangle$ and the two states $|1\rangle$ and $|-1\rangle$ would be degenerate. In addition, due to $\mathcal{T}^{2}=-1$, the states $|\pm 1\rangle$ and its time-reversal partner $\mathcal{T}|\pm 1\rangle$ are linearly independent. Hence, the four states $\{|1\rangle, \tilde{M}_{y}|1\rangle, \mathcal{T}|1\rangle, \mathcal{T}\tilde{M}_{y}|1\rangle\}$ must be degenerate at the same energy, forming a Dirac point.

Moreover, since $[\mathcal{P},C_{6z}]=0$, the chosen Bloch states at $A$ also can be the eigenstates of $C_{6z}$, which we denote as $|c_{6z},p\rangle$ with the $C_{6z}$ eigenvalue $c_{6z}\in \{\pm e^{i\pi/6}, \pm i, \pm e^{-i\pi/6}\}$. Based on Eq.~(\ref{aeq:c192_alg2}), one finds that
\begin{equation}
C_{6z}\tilde{M}_{y}|\pm i,\pm 1\rangle = -\tilde{M}_{y}C_{6z}^{5}|\pm i,\pm 1\rangle
=(\mp i)\tilde{M}_{y}|\pm i,\pm 1\rangle
\end{equation}
which indicates that $\tilde{M}_{y}|\pm i,\pm 1\rangle \Rightarrow |\mp i,\mp 1\rangle$.
Here, we choose the degenerate quartet states $\{|i, 1\rangle, \tilde{M}_{y}|i, 1\rangle, \mathcal{T}|i, 1\rangle, \mathcal{T}\tilde{M}_{y}|i, 1\rangle\}$ as our basis, so that the matrix representations of the generators can be expressed as
\begin{equation}
C_{6z} = i\sigma_{z}\otimes\sigma_{z}, \qquad
\tilde{M}_{y} = \sigma_{0}\otimes\sigma_{x}, \qquad
\mathcal{P} = \sigma_{0}\otimes\sigma_{z}, \qquad
\mathcal{T} = i\sigma_{y}\otimes\sigma_{0}\mathcal{K},
\end{equation}
and the symmetry constraints on Hamiltonian are
\begin{eqnarray}
C_{6z}{\cal H}(\bm{k})C_{6z}^{-1} & = & {\cal H}(R_{6z}\bm{k}), \label{aeq:c192_C6z} \\
\tilde{M}_{y}{\cal H}(\bm{k})\tilde{M}_{y}^{-1} & = & {\cal H}(k_{x},-k_{y},k_{z}), \label{aeq:c192_My}\\
\mathcal{P}{\cal H}(\bm{k})\mathcal{P}^{-1} & = & {\cal H}(-\bm{k}). \label{aeq:c192_P} \\
\mathcal{T}{\cal H}(\bm{k})\mathcal{T}^{-1} & = & {\cal H}(-\bm{k}). \label{aeq:c192_T}
\end{eqnarray}
One notes that from Eq. (\ref{aeq:c192_C6z}) together with $C_{3z}=-\sigma_{0}\otimes\sigma_{0}$, the Hamiltonian must satisfy $\mathcal{H}_{\text{eff}}(\bm{k})=\mathcal{H}_{\text{eff}}(R_{3z}\bm{k})$ which eliminates both linear and quadratic terms in $k_{x}$ and $k_{y}$, but the cubic term can be present. This indicates that the Dirac point would be a CDP.

Under the symmetry constraints in Eq.~(\ref{aeq:c192_C6z})-(\ref{aeq:c192_T}), the effective model expanded up to the third order is given as
\begin{eqnarray}
h_{11}^{192}(\bm{k}) & = & (c_{1}k_{+}^{3} + c_{2}k_{-}^{3})\sigma_{+} + \text{H.c.},\\
h_{12}^{192}(\bm{k}) & = & g(\bm{k})k_{z}\sigma_{x},
\end{eqnarray}
and
\begin{equation}
h_{22}^{192}(\bm{k})=h_{11}^{192*}(-\bm{k}).
\end{equation}
where $g(\bm{k})=\alpha_{1}+\alpha_{2}(k_{x}^{2}+k_{y}^{2})+\alpha_{3}k_{z}^{2}$.
This effective model confirms that the Dirac point is a CDP, with linear band splitting along $k_z$ and cubic splitting in the $k_x$-$k_y$ plane.

\subsubsection{The degeneracy in SG 176}
SG 176 has a fourfold degeneracy at the point $A$ ($0,0,\frac{1}{2}$). However, this is not an isolated Dirac point. It is on the intersection of three movable but unremovable Dirac nodal lines in the $k_{z}=\pi$ plane. The little group at $A$ contains three generators: $\tilde{C}_{6z}\equiv\left\{ C_{6z}^{+}\big|00\frac{1}{2}\right\}$, $\mathcal{P}\equiv\left\{\mathcal{P}\big|00\frac{1}{2}\right\}$ and $\mathcal{T}$. The following relations are satisfied at $A$:
\begin{equation}\label{aeq:c176_alg1}
\tilde{C}_{6z}^{6} = T_{003}\bar{E} = -e^{i2\pi (3\tilde{k}_{z})} = 1, \qquad
\mathcal{P}^{2} = 1
\end{equation}
and
\begin{equation}\label{aeq:c176_alg2}
\tilde{C}_{6z}\mathcal{P} = T_{001}\mathcal{P}\tilde{C}_{6z}
= e^{-i2\pi \tilde{k}_{z}}\mathcal{P}\tilde{C}_{6z}
= -\mathcal{P}\tilde{C}_{6z}.
\end{equation}
The Bloch states at $A$ can be chosen as the eigenstates of $\tilde{C}_{6z}$, which we denotes as $|c_{6z}\rangle$ with the $\tilde{C}_{6z}$ eigenvalue $c_{6z}\in \{\pm e^{i\pi/3}, \pm 1, \pm e^{-i\pi/3}\}$. Based on Eq.~(\ref{aeq:c176_alg2}), one finds that
\begin{equation}
\tilde{C}_{6z}\mathcal{P} |c_{6z}\rangle = -\mathcal{P}\tilde{C}_{6z}|c_{6z}\rangle
=-c_{6z}\mathcal{P}|c_{6z}\rangle,
\end{equation}
which indicates that $\mathcal{P} |c_{6z}\rangle \Rightarrow |-c_{6z}\rangle$ and the two states $|c_{6z}\rangle$ and $|-c_{6z}\rangle$ would be degenerate. In addition, due to $\mathcal{T}^{2}=-1$, the states $|c_{6z}\rangle$ and its time-reversal partner $\mathcal{T}|c_{6z}\rangle$ are linearly independent. Hence, the four states $\{|1\rangle, \mathcal{P}|1\rangle, \mathcal{T}|1\rangle, \mathcal{PT}|1\rangle\}$ must be degenerate at the same energy, while another four states $\{|e^{i\pi/3}\rangle, \mathcal{P}|e^{i\pi/3}\rangle, \mathcal{T}|e^{i\pi/3}\rangle, \mathcal{PT}|e^{i\pi/3}\rangle\}$ are also degenerate. In the following, we analyze the two cases separately. 

\paragraph{Basis: $\{|1\rangle, \mathcal{P}|1\rangle, \mathcal{T}|1\rangle, \mathcal{PT}|1\rangle\}$.} The matrix representations of the generators under this basis can be expressed as
\begin{equation}
\tilde{C}_{6z} = \sigma_{0}\otimes\sigma_{z}, \qquad
\mathcal{P} = \sigma_{0}\otimes\sigma_{x}, \qquad
\mathcal{T} = i\sigma_{y}\otimes\sigma_{0}\mathcal{K},
\end{equation}
and the symmetry constraints on Hamiltonian are
\begin{eqnarray}
\tilde{C}_{6z}{\cal H}(\bm{k})\tilde{C}_{6z}^{-1} & = & {\cal H}(R_{6z}\bm{k}), \label{aeq:c176_C6z} \\
\mathcal{P}{\cal H}(\bm{k})\mathcal{P}^{-1} & = & {\cal H}(-\bm{k}), \label{aeq:c176_P}\\
\mathcal{T}{\cal H}(\bm{k})\mathcal{T}^{-1} & = & {\cal H}(-\bm{k}). \label{aeq:c176_T}
\end{eqnarray}
One notes that from Eq. (\ref{aeq:c176_C6z}) together with $C_{6z}^2=\sigma_{0}\otimes\sigma_{0}$, the Hamiltonian must satisfy $\mathcal{H}_{\text{eff}}(\bm{k})=\mathcal{H}_{\text{eff}}(R_{3z}\bm{k})$ which eliminates both linear and quadratic terms in $k_{x}$ and $k_{y}$, but the cubic term can be present. 

Under the symmetry constraint in Eq.~(\ref{aeq:c176_C6z})-(\ref{aeq:c176_T}), the effective model expanded up to the third order is given as
\begin{eqnarray}
h_{11}^{176}(\bm{k}) & = & f(\bm{k})k_{z}\sigma_{z} + (\alpha_{1}k_{+}^{3}+\alpha_{1}^{*}k_{-}^{3})\sigma_{y},\\
h_{12}^{176}(\bm{k}) & = & g(\bm{k})k_{z}\sigma_{z},
\end{eqnarray}
and
\begin{equation}
h_{22}^{176}(\bm{k})=h_{11}^{176*}(-\bm{k}).
\end{equation}
where $f(\bm{k})=c_{1}+c_{2}(k_{x}^{2}+k_{y}^{2})+c_{3}k_{z}^{2}$ and $g(\bm{k})=\alpha_{2}+\alpha_{3}(k_{x}^{2}+k_{y}^{2})+\alpha_{4}k_{z}^{2}$.
This result shows a linear band splitting along $k_z$ and cubic splitting in the $k_x$-$k_y$ plane.

However, in the $k_{z}=0$ plane,
the dispersion is given by
\begin{equation}\label{aeq:c176_DNL}
E = \pm (\alpha_{1}k_{+}^{3}+\alpha_{1}^{*}k_{-}^{3}).
\end{equation}
Here, due to screw rotation $\tilde{C}_{6z}$, each band at the $k_z=0$ plane (notice that here $k_z$ is measured from the  $A$ point) is at least doubly degenerate.
From this dispersion (\ref{aeq:c176_DNL}), one finds that a fourfold degeneracy appears at the path defined by
\begin{equation}
\alpha_{1}k_{+}^{3}+\alpha_{1}^{*}k_{-}^{3} = 0,
\end{equation}
which is satisfied when $\cos(3\phi+\theta)=0$ with $\theta=\arg{(\alpha_{1})}$ and $\phi=\arg{(k_x+i k_y)}$. The condition  $\cos(3\phi+\theta)=0$ gives three Dirac nodal lines at the plane and the three lines intersect at the $A$ point.

\paragraph{Basis: $\{|e^{i\pi/3}\rangle, \mathcal{P}|e^{i\pi/3}\rangle, \mathcal{T}|e^{i\pi/3}\rangle, \mathcal{PT}|e^{i\pi/3}\rangle\}$.} The matrix representations of the generators can be expressed as
\begin{equation}
\begin{aligned}
\tilde{C}_{6z} = e^{i\frac{\pi}{3}\sigma_{z}}\otimes\sigma_{z}, \qquad
\mathcal{P} = \sigma_{0}\otimes\sigma_{x}, \qquad
\mathcal{T} = i\sigma_{y}\otimes\sigma_{0}\mathcal{K}.
\end{aligned}
\end{equation}
Together with the symmetry constraints in Eq.~(\ref{aeq:c184_C6z})-(\ref{aeq:c184_T}), the effective model expanded up to the third order is given as
\begin{eqnarray}
h_{11}^{176}(\bm{k}) & = & f(\bm{k})k_{z}\sigma_{z} + (\alpha_{1}k_{+}^{3}+\alpha_{1}^{*}k_{-}^{3})\sigma_{y},\\
h_{12}^{176}(\bm{k}) & = & \alpha_{2} k_{-}^{2}k_{z}\sigma_{z},
\end{eqnarray}
and
\begin{equation}
h_{22}^{176}(\bm{k})=h_{11}^{176*}(-\bm{k}).
\end{equation}
where $f(\bm{k})=c_{1}+c_{2}(k_{x}^{2}+k_{y}^{2})+c_{3}k_{z}^{2}$.
Similarly, one can check that this point is also located at the intersection of  three movable but unremovable Dirac nodal lines in the $k_z=0$ plane.


\section{Lattice model of SG 92}\label{ASec_TBmodel}

\begin{figure}[h]
	\includegraphics[width=8.5cm]{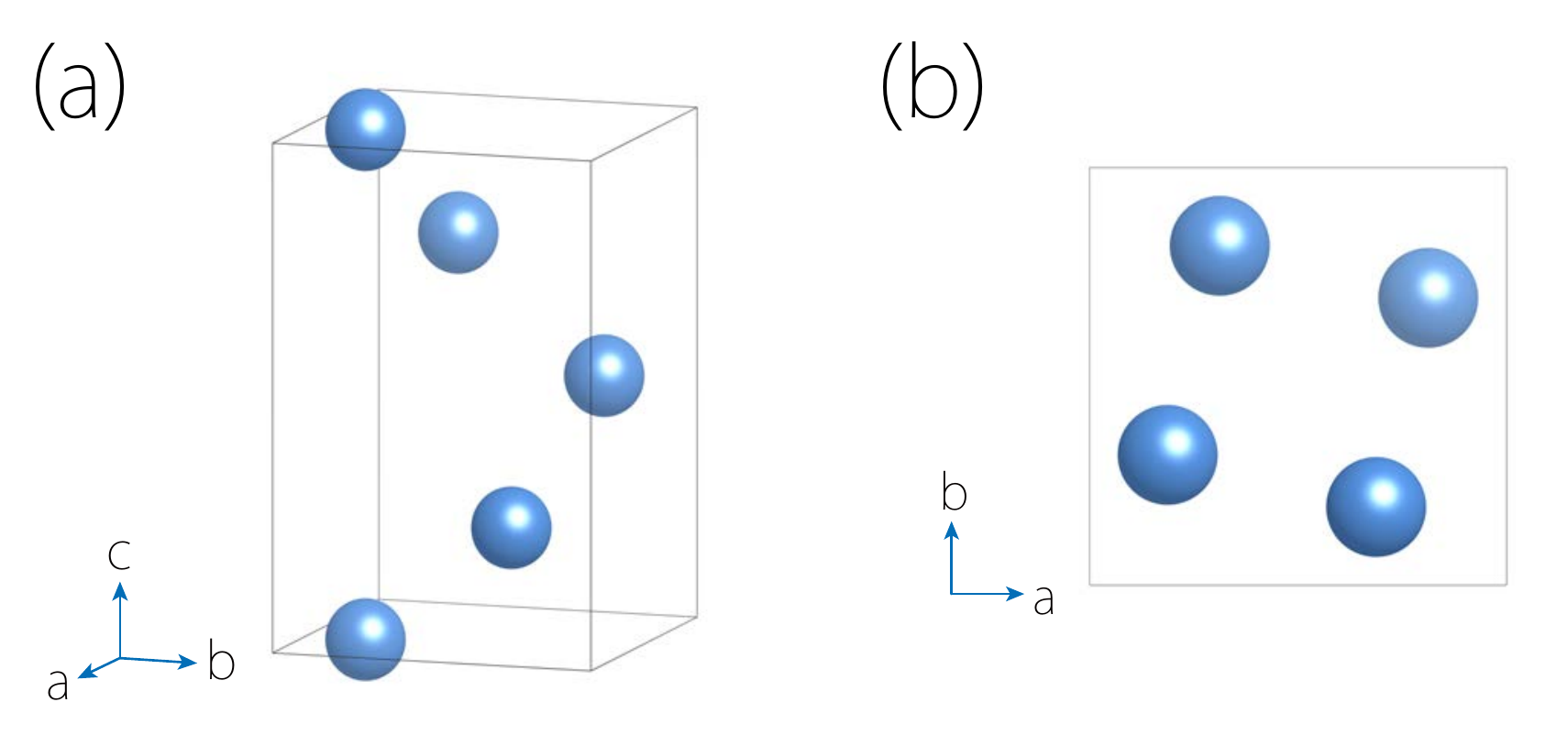}
	\caption{Lattice model of SG 92. (a) Perspective view and (b) top view of the primitive unit cell of the lattice. The origin has been shifted by $\bm{t}_{0}=(\frac{1}{2},0,0)$ compared with the standard setting. \label{fig_af1}}
\end{figure}

To study the chiral QDP, we have constructed a minimal lattice model for SG 92. The atomic lattice is shown in Fig.~\ref{fig_af1}. In this model,
each unit cell contains four sites denoted by the blue balls and each site has an $s$-like orbital with two spin states, such that the lattice model with SOC considered have eight bands.
The model is constrained by the following symmetry generators for the SG: 
$\tilde{C}_{4z}\equiv\{C_{4z}\big|00\frac{1}{4}\}$, $\tilde{C}_{2x}\equiv\left\{ C_{2x}\big|\frac{1}{2}\frac{1}{2}\frac{3}{4}\right\}$,
and ${\cal T}$.
We order the eight basis states in the unit cell as $(|A_{1},\uparrow\rangle,\ |A_{1},\downarrow\rangle,\ |A_{2},\uparrow\rangle,\ |A_{2},\downarrow\rangle,\ |A_{3},\uparrow\rangle,\ |A_{3},\downarrow\rangle,\ |A_{4},\uparrow\rangle,\ |A_{4},\downarrow\rangle)$, where
$A_{i}$ ($i=1,2,3,4$) denote the four atoms occupying the $4a$ Wyckoff positions $\left\{ (-\frac{5}{16},\frac{3}{16},0),\ (-\frac{3}{16},-\frac{5}{16},\frac{1}{4}),\ (\frac{5}{16},-\frac{3}{16},\frac{1}{2}),\ (\frac{3}{16},\frac{5}{16},\frac{3}{4})\right\} $. In this basis, the
symmetry operators are represented as
\begin{equation}
\begin{aligned}
\tilde{C}_{4z} = \left[
\begin{array}{cccc}
0 & 1 & 0 & 0\\
0 & 0 & 1 & 0\\
0 & 0 & 0 & 1\\
1 & 0 & 0 & 0
\end{array}
\right] \otimes e^{i\frac{\pi}{4}\sigma_{z}}, \qquad
\tilde{C}_{2x} = \left[
\begin{array}{cccc}
0 & 0 & 0 & 1\\
0 & 0 & 1 & 0\\
0 & 1 & 0 & 0\\
1 & 0 & 0 & 0
\end{array}
\right] \otimes i\sigma_{x}, \qquad
{\cal T} = -i\sigma_{y}{\cal K}.
\end{aligned}
\end{equation}
Then the symmetry allowed minimal lattice model up to the second-neighbor hopping can be obtained as
\begin{equation}
\mathcal{H}_{92}^\text{TB}=\varepsilon_{0}+\left[
\begin{array}{cccc}
0 & h_{12}(\bm{k}) & h_{13}(\bm{k}) & h_{14}(\bm{k})\\
h_{12}^{\dagger}(\bm{k}) & 0 & h_{23}(\bm{k}) & h_{24}(\bm{k})\\
h_{13}^{\dagger}(\bm{k}) & h_{23}^{\dagger}(\bm{k}) & 0 & h_{34}(\bm{k})\\
h_{14}^{\dagger}(\bm{k}) & h_{24}^{\dagger}(\bm{k}) & h_{34}^{\dagger}(\bm{k}) & 0
\end{array}
\right],
\end{equation}
with
\begin{equation}
\begin{aligned}
h_{12}(\bm{k})&=\left(a_{0}\cos\frac{k_{y}}{2}\sigma_{0}+ia_{1}\cos\frac{k_{y}}{2}\sigma_{x}
-a_{2}\sin\frac{k_{y}}{2}\sigma_{y}+ia_{3}\cos\frac{k_{y}}{2}\sigma_{z}\right)e^{i(\frac{k_{x}}{8}+\frac{k_{z}}{4})}, \\
h_{13}(\bm{k})&=\left(b_{0}\cos\frac{k_{z}}{2}\sigma_{0}-ib_{1}\cos\frac{k_{z}}{2}\sigma_{x}
-ib_{1}\cos\frac{k_{z}}{2}\sigma_{y}-b_{2}\sin\frac{k_{z}}{2}\sigma_{z}\right)e^{i(-\frac{3k_{x}}{8}-\frac{3k_{y}}{8})}, \\
h_{14}(\bm{k})&=\left(a_{0}\cos\frac{k_{x}}{2}\sigma_{0}-a_{2}\sin\frac{k_{x}}{2}\sigma_{x}
+ia_{1}\cos\frac{k_{x}}{2}\sigma_{y}-ia_{3}\cos\frac{k_{x}}{2}\sigma_{z}\right)e^{i(\frac{k_{y}}{8}-\frac{k_{z}}{4})}, \\
h_{23}(\bm{k})&=\left(a_{0}\cos\frac{k_{x}}{2}\sigma_{0}-a_{2}\sin\frac{k_{x}}{2}\sigma_{x}
+ia_{1}\cos\frac{k_{x}}{2}\sigma_{y}+ia_{3}\cos\frac{k_{x}}{2}\sigma_{z}\right)e^{i(\frac{k_{y}}{8}+\frac{k_{z}}{4})}, \\
h_{24}(\bm{k})&=\left(b_{0}\cos\frac{k_{z}}{2}\sigma_{0}+ib_{1}\cos\frac{k_{z}}{2}\sigma_{x}
-ib_{1}\cos\frac{k_{z}}{2}\sigma_{y}-b_{2}\sin\frac{k_{z}}{2}\sigma_{z}\right)e^{i(\frac{3k_{x}}{8}-\frac{3k_{y}}{8})}, \\
h_{34}(\bm{k})&=\left(a_{0}\cos\frac{k_{y}}{2}\sigma_{0}-ia_{1}\cos\frac{k_{y}}{2}\sigma_{x}
-a_{2}\sin\frac{k_{y}}{2}\sigma_{y}+ia_{3}\cos\frac{k_{y}}{2}\sigma_{z}\right)e^{i(-\frac{k_{x}}{8}+\frac{k_{z}}{4})},
\end{aligned}
\end{equation}
where the $\sigma$'s are the Pauli matrices and $\sigma_{0}$ denotes the $2\times2$ identity matrix. The coefficients $\varepsilon_{0}$, $a_{i}$ and $b_{i}$ are real model parameters. For the results shown in the main text, we have taken the following parameter values: $a_{0}=0.8$, $a_{2}=0.25$, $a_{3}=-0.05$, $b_{1} = -1.0$, and the other parameters are set to be zero.

\section{Solution of Chiral Landau bands}\label{ASec_Landau}
The chiral QDP in SG~92 leads to four chiral Landau bands since it has a topological charge of $\pm 4$, as shown in Sec.~V.B of the main text. In this section, we attempt an analytic solution of these Landau bands. It is challenging to directly solve the effective model derived for SG~92 [Eq.~(8-10) in the main text]. To proceed, we make a simplification by dropping the two non-diagonal blocks in the model. This does not alter the essential physics, because the topological charge is unaffected and the dispersion type is also maintained.
This simplified model for the chiral QDP reads:
\begin{equation}
H(\bm{k})= \left[
\begin{array}{cc}
\lambda k_{z}\sigma_{z} + \alpha k_{+}^{2} \sigma_{+} + \alpha^{*} k_{-}^{2} \sigma_{-} & 0 \\
0 & -\lambda k_{z}\sigma_{z} + \alpha^{*} k_{-}^{2} \sigma_{+} + \alpha k_{+}^{2} \sigma_{-}
\end{array}
\right].
\end{equation}

Under an external magnetic field ($\bm{B}=-B\hat{z}$), we make the usual Peierls substitution $\bm{k} \rightarrow \bm{\Pi} = \bm{k}+\frac{e}{\hbar c}\bm{A}$ and choose the Landau gauge ($\bm{A}=-Bx\hat{y}$). Since $[\Pi_{x},\Pi_{y}] = i\hbar/\ell_{B}^{2}$ with $\ell_{B} = \sqrt{\hbar c/eB}$ the magnetic length, it is convenient to introduce the ladder operators, $a=\frac{\ell_{B}}{\sqrt{2}\hbar}(\Pi_{x}+i\Pi_{y})$ and $a^{\dagger}=\frac{\ell_{B}}{\sqrt{2}\hbar}(\Pi_{x}-i\Pi_{y})$, obeying $[a,a^{\dagger}]=1$. In terms of the ladder operators, the Hamiltonian becomes
\begin{equation}
H=\left[
\begin{array}{cccc}
\lambda k_{z} & \alpha \omega_{c} a^{2} & 0 & 0\\
\alpha^{*} \omega_{c} a^{\dagger 2} & -\lambda k_{z} & 0 & 0 \\
0 & 0 & -\lambda k_{z} & \alpha^{*} \omega_{c} a^{\dagger 2} \\
0 & 0 & \alpha \omega_{c} a^{2} & \lambda k_{z}
\end{array}
\right],
\end{equation}
where $\omega_{c} = \sqrt{2}\hbar/\ell_{B}$. The harmonic oscillator eigenstates are given by $|n\rangle=\frac{1}{\sqrt{n!}}(a^{\dagger})^{n}|0\rangle$, with $a|0\rangle=0$, $a|n\rangle=\sqrt{n}|n-1\rangle$, $a^{\dagger}|n\rangle=\sqrt{n+1}|n+1\rangle$. Then the eigenvalues and eigenstates of $H$ are obtained by solving the equation $H\psi_{n}=\varepsilon \psi_{n}$. One can find that the eigenstates $\psi_{n}$ take the following form
\begin{equation}
\psi_{n}=\begin{cases}
(c_{1}|n-2\rangle,\ c_{2}|n\rangle,\ c_{3}|n-1\rangle,\ c_{4}|n-3\rangle)^{T}, & n\geq3\\
(c_{1}|0\rangle,\ c_{2}|2\rangle,\ c_{3}|1\rangle,\ 0)^{T}, & n=2\\
(0,\ c_{2}|1\rangle,\ c_{3}|0\rangle,\ 0)^{T}, & n=1\\
(0,\ c_{2}|0\rangle,\ 0,\ 0)^{T}, & n=0
\end{cases}
\end{equation}
where $c_{1,2,3,4}$ are the parameters to be determined. The eigenvalues $\varepsilon_{n}$ are obtained by solving $|\varepsilon_{n}-Q_{n}|=0$ with
\begin{eqnarray}
Q_{n \geq 3} & = & \left[
\begin{array}{cccc}
\lambda k_{z} & \alpha \omega_{c} \sqrt{n(n-1)} & 0 & 0\\
\alpha^{*} \omega_{c} \sqrt{n(n-1)} & -\lambda k_{z} & 0 & 0 \\
0 & 0 & -\lambda k_{z} & \alpha^{*} \omega_{c} \sqrt{(n-1)(n-2)} \\
0 & 0 & \alpha \omega_{c}\sqrt{(n-1)(n-2)} & \lambda k_{z}
\end{array}
\right], \\
Q_{n = 2} & = & \left[
\begin{array}{ccc}
\lambda k_{z} & \alpha \omega_{c} \sqrt{2} & 0 \\
\alpha^{*} \omega_{c} \sqrt{2} & -\lambda k_{z} & 0 \\
0 & 0 & -\lambda k_{z}
\end{array}
\right], \\
Q_{n=1} & = & -\lambda k_{z} \sigma_{0}, \\
Q_{n=0} & = & -\lambda k_{z}
\end{eqnarray}

which gives
\begin{eqnarray}
\varepsilon_{n\geq3} & = & 
\pm \sqrt{\lambda^{2}k_{z}^{2}+n(n-1)|\alpha|^{2}\omega_{c}^{2}}, \quad
\pm \sqrt{\lambda^{2}k_{z}^{2}+(n-1)(n-2)|\alpha|^{2}\omega_{c}^{2}} \\
\varepsilon_{2} & = & 
\pm \sqrt{\lambda^{2}k_{z}^{2}+2|\alpha|^{2}\omega_{c}^{2}}, \quad
-\lambda k_{z} \\
\varepsilon_{1} & = & -\lambda k_{z}, \quad -\lambda k_{z} \\
\varepsilon_{0} & = & -\lambda k_{z}.
\end{eqnarray}
It is straightforward to see that there are four chiral Landau bands: one from $n=2$, two from $n=1$, and one from $n=0$. 

%



%